\documentclass{aa}  
\usepackage{natbib}
\usepackage{graphicx}
\usepackage{txfonts}
\usepackage{caption}
\usepackage{subcaption}
\usepackage{amsmath,amsfonts,bm}
\usepackage{comment}
\usepackage[dvipsnames]{xcolor}   

\usepackage{soul}
\usepackage{xcolor}

\begin{document}

   \title{Icarus 3.0: Dynamic Heliosphere Modelling}
   \author{T. Baratashvili \inst{1}, B. Popescu Braileanu\inst{1}, F. Bacchini\inst{1,2}, R. Keppens\inst{1}, S. Poedts \inst{1,3}
          }

   \institute{Department of Mathematics, Centre for mathematical Plasma Astrophysics, 
             KU Leuven, 3001 Leuven, Belgium\\
             \email{tinatin.baratashvili@kuleuven.be}
             \and
             Royal Belgian Institute for Space Aeronomy, Solar-Terrestrial Centre of Excellence, Ringlaan 3, 1180 Uccle, Belgium
             \and
             Institute of Physics, University of Maria Curie-Sk{\l}odowska, 
             PL-20 031 Lublin, Poland
             }

   \date{Received: \today}

  \abstract
   {Space weather predictions are necessary to avoid damage caused by intense geomagnetic storms. Such strong storms are usually caused by a co-rotating interaction region (CIR) passing at Earth or by the arrival of strong coronal mass ejections (CMEs). To mitigate the damage, the effect of propagating CMEs in the solar wind must be estimated accurately at Earth and other locations. Modelling solar wind accurately is crucial for space weather predictions, as it is the medium for CME propagation.  
   }
   {The \texttt{Icarus} heliospheric modelling tool is upgraded to handle dynamic inner heliospheric driving instead of using steady boundary conditions. The ideal magnetohydrodynamic (MHD) solver and the automated grid-adaptivity are adjusted to the latest \texttt{MPI-AMRVAC} version. This new combination allows us to model the solar heliosphere more accurately.
   }
   {The inner boundary conditions, prescribed at 0.1~AU for the heliospheric model, are updated time-dependently throughout the simulation. The coronal model ($r<0.1 $ AU) is computed repeatedly for selected magnetograms, and the $r=0.1$ AU radial boundary prescription is provided to the heliospheric modelling tool. The particle sampling within \texttt{MPI-AMRVAC} is extended to handle stretched spherical grid information. It is well suited for tracing solar wind plasma conditions at the locations of planets and satellites in the heliosphere. 
   }
   {The solar wind obtained in the simulation is dynamic and shows significant variations throughout the evolution. When comparing the results with the observations, the dynamic solar wind results are more accurate than previous results obtained with purely steady boundary driving. The CMEs propagated through the dynamic solar wind background produce more similar signatures in the time-series data than in the steady solar wind. 
   }
   {Dynamic boundary driving in \texttt{Icarus} results in a more self-consistent solar wind evolution in the inner heliosphere. The upgraded particle sampling allows for a very versatile sampling of the solution at the spatio-temporally varying locations of satellites. The obtained space weather modelling tool for
   dynamic solar wind and CME simulations is better suited for space weather forecasting than a steady solar wind model.
   }


\titlerunning{Icarus 3.0}
\authorrunning{Baratashvili et al.}
\maketitle
\section{Introduction}
The solar wind accelerates and becomes supersonic in the solar corona. The solar wind is observed to have a bi-modal structure, which originates from different sources. The origin of the slow solar wind is still under debate among solar physicists and is believed to be coming from the coronal hole boundaries \citep{antiochos2013}, but the fast solar wind comes from the coronal hole regions. These regions are not fixed on the solar surface and change over time. Therefore, the solar wind in the heliosphere is also not static and undergoes significant variations when the activity on the solar surface changes. The solar wind is the medium where strong solar eruptions, such as Coronal Mass Ejections (CMEs), propagate. The magnetized ejecta interact with the solar wind's magnetic field and experience deformation during their propagation. Strong CMEs often cause strong geomagnetic storms on Earth and cause significant damage to power systems and telecommunications. Space weather is a branch of physics focusing on the disturbances caused by solar activity in the heliosphere. Space weather modelling is essential for mitigating potential damage on operating spacecraft, Earth, and other locations in the heliosphere. Most space weather modelling tools use steady boundary conditions for the heliospheric model at 0.1~AU for the predictions. This implies selecting a magnetogram corresponding to the date of interest, performing the coronal computations, and starting the heliosphere modelling with this single set of boundary information, i.e. without further updating the input data, even though the simulations run for periods of 10-25 days. Estimating the solar wind conditions near Earth and at the locations of selected spacecraft is important \citep{Baker2009}, and the steady solar wind approximation is usually justified under solar minimum conditions. During these periods, the wind is more homogeneous, the conditions do not change rapidly, and no significant details are missed by keeping the background wind steady.
Moreover, even under these quiet circumstances, the predictions are usually not done for periods longer than a week, as the input magnetogram becomes ``outdated'' to generate the conditions in the outer heliosphere or even beyond the orbit of Mars. The ambient solar wind plays a critical role in modelling the propagation of CMEs and predicting their trajectory, speed, and arrival time at a given location \citep{Manchester2008, odstrcil2009, lionello2013}. \cite{lee2013} showed that a more accurate solar wind model yields better predictions of CME arrival times at 1~AU. Steady heliospheric driving imposes significant limitations for modelling the conditions near Jupiter and for solar energetic particle (SEP) studies, as in both cases, updating the solar wind during the simulation is crucial to model the heliosphere at larger distances accurately.

A few models have attempted time-dependent solar wind modelling in the past. \cite{Linker2016} modelled solar wind time-dependently with the Wang-Sheeley-Arge (WSA) \citep{arge2003} empirical coronal model and Magnetohydrodynamic Algorithm outside a Sphere (MAS) \citep{riley2012} heliospheric model for five years between September 27, 2003, and September 27, 2008, with the cadence of 1 day.  \citet{kim2014} and \cite{jackson2015} introduced time-dependent heliospheric simulations using interplanetary scintillation data as time-dependent boundary conditions to solve the magnetohydrodynamic (MHD) equations in the heliosphere. \cite{hayashi2012} developed a time-dependent MHD model that starts at $50\;$R$_\odot$ and uses Wilcox Solar Observatory (WSO) magnetograms to obtain magnetic field boundary conditions at the inner boundary. It does so by extrapolating the radial magnetic field at the $2.5\;$R$_\odot$ source surface map of the Potential Field Source Surface (PFSS) and using solar wind speed data derived from interplanetary scintillation (IPS) observations at Nagoya University, Japan, for driving the inner boundary. The simulation was performed for seven months in 1991, and the cadence of updating magnetograms was one solar rotation. \cite{Merkin2016} demonstrated the upgrade of the Lyon-Fedder-Mobarry MHD code \citep{Merkin2011,Pahud2012} to include an inner boundary that is evolving in time. The daily magnetograms were obtained from the Air ForceData Assimilative Photospheric Flux Transport (ADAPT) model in this approach. The inner boundary conditions were calculated with the WSA coronal model using time-evolving magnetograms instead of the usual steady boundary treatment. 

In this work, the newly developed heliospheric model \texttt{Icarus} \citep{Verbeke2022,Baratashvili2022} was upgraded first to the latest updates of its numerical framework \texttt{MPI-AMRVAC~3.0} \citep{Keppens2023}. The code is maintained within the \texttt{MPI-AMRVAC} GitHub repository\footnote{\url{https://github.com/amrvac/amrvac/tree/master/tests/mhd/icarus}}. The built-in functionalities of \texttt{MPI-AMRVAC~3.0} were used to the full extent by exploiting its options to obtain results in the reference frame co-rotating with the Sun, to trace the planets and satellites in the domain and to sample solar wind data at these varying locations. The treatment of the inner boundary was also entirely re-implemented to benefit from the latest developments in \texttt{MPI-AMRVAC}. Moreover, time-dependent inner boundary conditions are introduced to evolve the plasma conditions at the inner heliospheric boundary based on the most up-to-date magnetogram information varying during the simulation. This upgrade is a significant new development for the heliospheric model. It expands the scope of studies that can be conducted with \texttt{Icarus}, particularly beyond 1~AU. It also enables more precise solar wind modelling during solar maxima, which provides a more accurate ambient solar wind for CME propagation. This upgrade is a crucial step towards continuous live space weather forecasting. 
Further, the implementation of the boundary conditions was modified to enable the injection of the CMEs superposed on the varying solar wind at the inner heliospheric boundary. All these developments led to the implementation of a new default boundary condition type called {\fontfamily{pcr}\selectfont bc\_icarus}, which treats the inner boundary according to those required for \texttt{Icarus} heliosphere simulations. 

The paper is organized as follows: Section~\ref{icarus_setup} introduces the updates in the heliospheric model compatible with the latest \texttt{MPI-AMRVAC~3.0} version. The upgraded dynamic boundary conditions are described in Section~\ref{dynamic_bcs}. The solar wind modelled for 42 days in September-October 2019 is considered in Section~\ref{2019case}, and Section~\ref{2015case} demonstrates the results for the solar wind and five consecutive CMEs in June-July 2015. The conclusions and prospects are summarized in Section~\ref{conclusions}.

\section{Icarus 3.0: Numerical Setup}
\label{icarus_setup}
\texttt{Icarus} was developed as an alternative to the EUropean Heliospheric FORecasting Information Asset (EUHFORIA) heliospheric model \citep{Pomoell2018}, covering the same physical domain as EUHFORIA. However, in \texttt{Icarus}, the equations are solved in a frame co-rotating with the Sun. At the same time, EUHFORIA uses Heliocentric Earth Equatorial (HEEQ) coordinates (in which the longitude of the Earth always remains zero). As a result, the relaxed wind in \texttt{Icarus} is steady in the co-rotating frame while the implemented planets and satellites move along their orbits. Using a co-rotating frame requires adding source terms in the momentum and energy equations representing the centrifugal and Coriolis forces. These terms are right-hand-side sources in Equations~\eqref{eq:icarus_momentum} and \eqref{eq:icarus_energy}.

\begin{align}\label{eq:mhd}
    \frac{\partial \rho}{\partial t}+ \nabla \cdot (\rho \mathbf{v}) & =0,\\
    \frac{\partial (\rho\mathbf{v})}{\partial t} + \nabla \cdot \bigg(\rho\mathbf{v}\mathbf{v}+p_{tot}\mathbf{I}-\mathbf{B}\mathbf{B} \bigg) - \rho \mathbf{g} &= \mathbf{F}, \label{eq:icarus_momentum}\\
    \frac{\partial e}{\partial t} + \nabla \cdot \bigg( e\mathbf{v} + p_{tot}\mathbf{v} - \mathbf{B}(\mathbf{B}\cdot \mathbf{v}) \bigg) - \rho\mathbf{v} \cdot \mathbf{g}  & = \mathbf{v} \cdot \mathbf{F} , \label{eq:icarus_energy} \\
     \frac{\partial \mathbf{B}}{\partial t} + \nabla\cdot \bigg(\mathbf{v}\mathbf{B} - \mathbf{B}\mathbf{v}\bigg)&=0,\label{eq:faraday} \\
    \nabla \cdot \mathbf{B} &=0,
\end{align}
where
\begin{equation}
\quad p_{tot} = (\gamma-1)\bigg(e-\rho \frac{\mathbf{v}^2}{2}-\frac{\mathbf{B}^2}{2}\bigg) + \frac{\mathbf{B}^2}{2},
\end{equation}
and $\rho$, $\mathbf{v}$, $p_{tot}$, $\mathbf{B}$, and $e$
are the mass density, the velocity vector field, the total pressure (i.e.\ the sum of the thermal and magnetic pressure) of the plasma, the magnetic field vector, and the total energy density, respectively. We refer to \cite{Porth2014} for more details about the implementation into the \texttt{MPI-AMRVAC} framework. We consider the magnetic permeability $\mu_0$ to be 1 in nondimensional units. The gravitational acceleration $\mathbf{g}$ is given by $({GM_{\odot}}/{r^2}\mathbf{e_r})$, with $G$ the gravitational constant and $M_{\odot}$ the mass of the Sun. The adiabatic index $\gamma$ is 1.5, similar to \cite{Pomoell2018} and \cite{Odstrcil2004}. This reduced index models additional heating in the simplest way and yields solar wind acceleration throughout the inner heliosphere (see, e.g.,  \cite{Pomoell2012}). Finally, the source term $\mathbf{F}$ is given by $\rho (\mathbf{\Omega} \times \mathbf{r}) \times \mathbf{\Omega} + 2 \rho (\mathbf{v}\times\mathbf{\Omega})$, corresponding to the centrifugal and Coriolis forces, respectively. Here, $\Omega$ (the magnitude of the vector $\mathbf{\Omega}$) is the rotation rate of the Sun at the equator ($2.97\cdot 10^{-6}\;$rad/s).

In release \texttt{MPI-AMRVAC 3.0} \citep{Keppens2023}, built-in functionalities include a precoded way to handle the co-rotating frame with the Sun in the MHD module and the possibility to do flexible particle tracing or sampling in the domain. Therefore, implementing the centrifugal and Coriolis force terms was removed from the \texttt{Icarus} user file, and the built-in co-rotating frame functionality was activated. The option introduces the same terms in the MHD equations, and the co-rotating frame is thus achieved with this option. The rotation rate is passed from the input parameter file to account for the rotation rate implemented before. Generic handling of a co-rotating frame is now available for hydrodynamic and MHD applications in cylindrical and spherical coordinates in \texttt{MPI-AMRVAC~3.0}. 

\begin{figure}[hbt!]
\centering
    \includegraphics[width=0.5\textwidth]{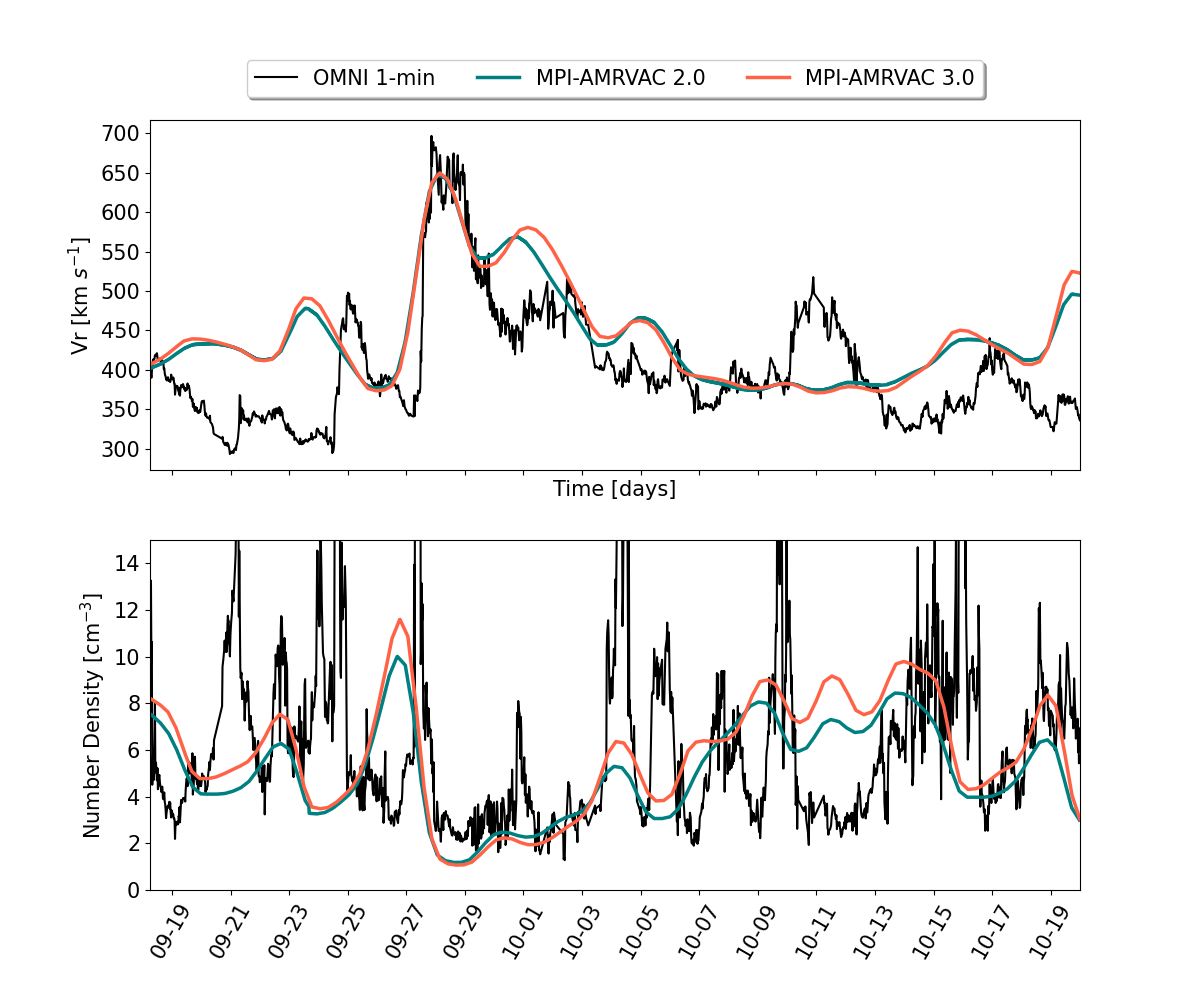}
  \caption{Radial velocity (upper panel) and number density (lower panel) time series at Earth simulated with \texttt{Icarus} in \texttt{MPI-AMRVAC~2.0} and \texttt{AMRVAC~3.0} versions. OMNI 1-min data are plotted in black with a 20-minute filter.}\label{fig:amrvac2_amrvac3}
\end{figure}
Another update concerns how we want to accumulate solar wind data tracing the planets and satellites in the heliosphere. Initially, \texttt{Icarus} realized this by a user-specific means to interpolate data at desired locations within the generic interface meant to add source terms to the MHD equations. In this case, the data points along the trajectories of different satellites were interpolated from the computational grid and saved as time-series files for each tracing object. The recently upgraded particle module of \texttt{MPI-AMRVAC~3.0} \citep{Keppens2023,bacchini2024} provides the same functionality of sampling data at spatiotemporally prescribed locations but had to be significantly adjusted to handle also radially stretched, hierarchical grid data widely used in \texttt{Icarus}. With this setting, the particles are introduced as moving sampling location points and do not have mass or velocity.

Additionally, the location of each satellite is now interpolated from the obtained trajectory files. The data for each particle is stored as an individual data file that can be manipulated for analysis. When the satellites are traced using the particle sampling method, the default output format is converted to the standard EUHFORIA output format. The two versions were compared for the same input boundary file generated with the WSA model. The GONG magnetogram corresponding to 2019.09.18 06:14:00 was taken. The coronal model configuration file was modified the same way as in \cite{Wijsen2021}. Figure~\ref{fig:amrvac2_amrvac3} shows the comparison for the time series obtained from the \texttt{Icarus} simulation with the \texttt{MPI-AMRVAC}~2.0 and 3.0 versions in teal and orange, respectively. The black curve represents the OMNI 1~min data. The upper panel shows the velocity values and the lower panel shows the number density values at the Earth's location. Here, we can see slight differences between the two curves due to the interpolation differences between the locations when updating the trajectory of the satellites.

\subsection{Updated Boundary Conditions}
\begin{figure}[hbt!]
    \includegraphics[width=0.5\textwidth]{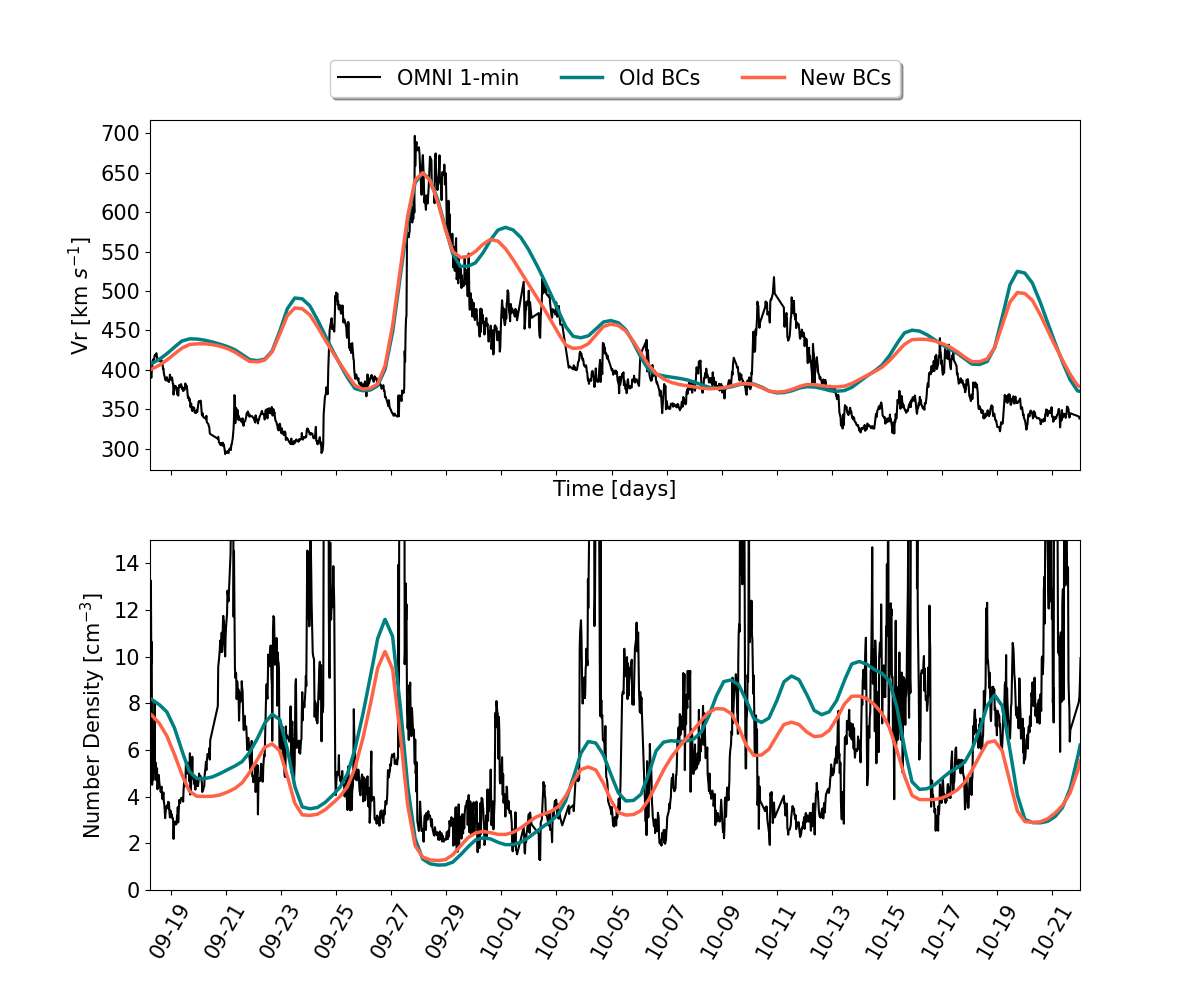}
  \caption{Radial velocity (upper panel) and number density (lower panel) at Earth simulated with \texttt{Icarus 3.0} with old and new boundary conditions. OMNI 1-min data are plotted in black with a 20-minute filter.} \label{fig:amrvac3_old_new}
\end{figure}

\begin{figure*}[hbt!]
\centering
    \begin{subfigure}[b]{0.33\textwidth}
         \centering
         \includegraphics[width=\textwidth]{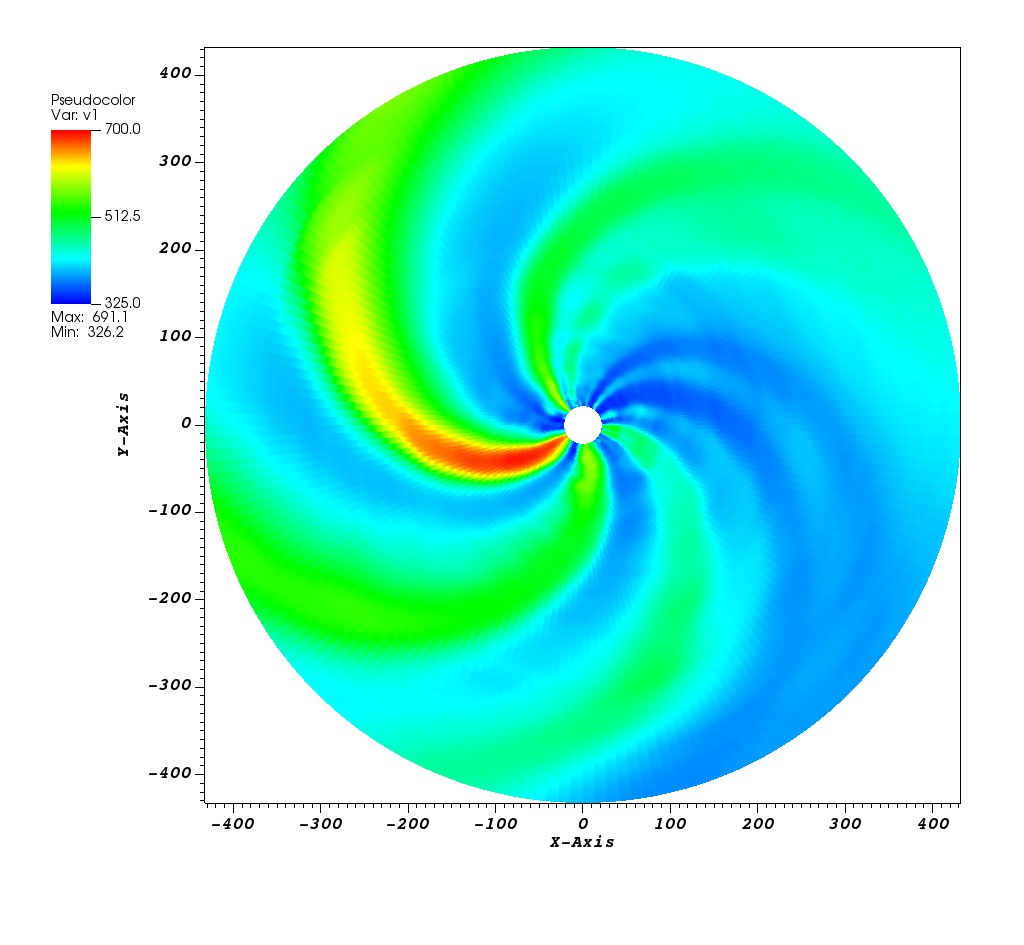}

     \end{subfigure}
     \hfill
          \begin{subfigure}[b]{0.33\textwidth}
         \centering
         \includegraphics[width=\textwidth]{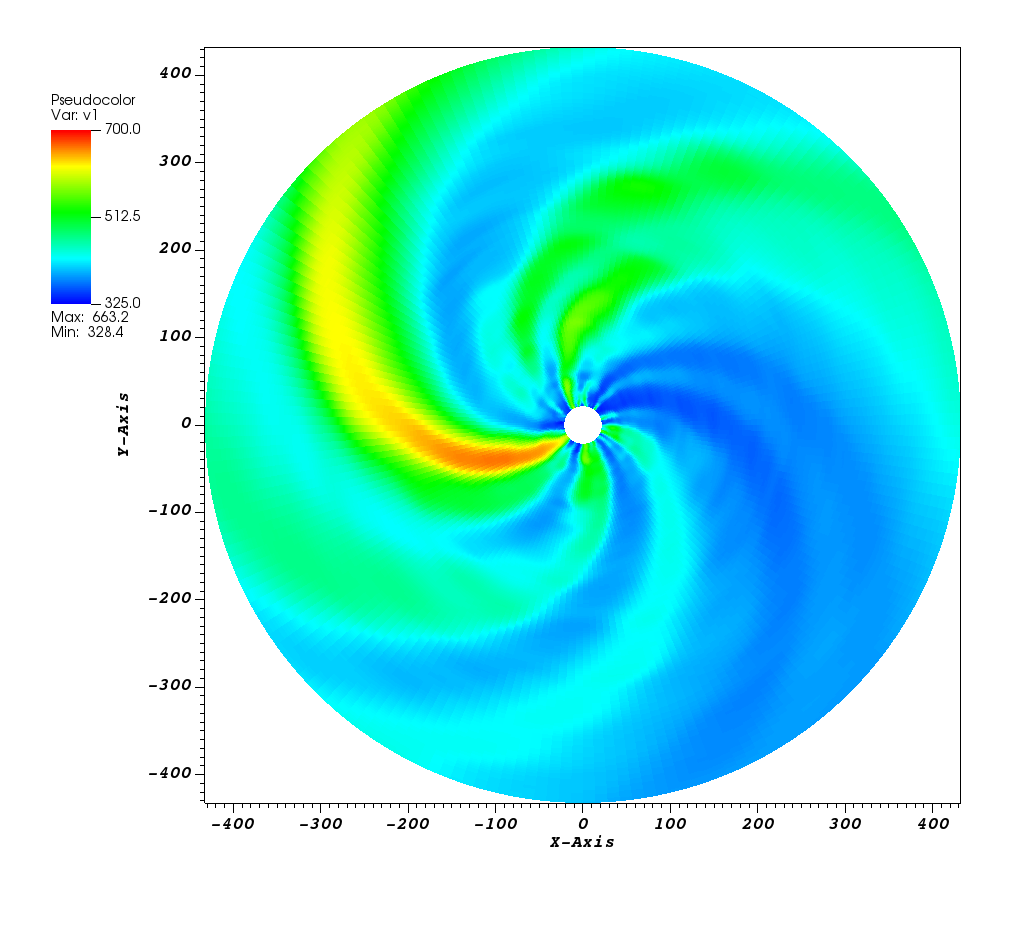}

     \end{subfigure}
     \hfill
          \begin{subfigure}[b]{0.33\textwidth}
         \centering
         \includegraphics[width=\textwidth]{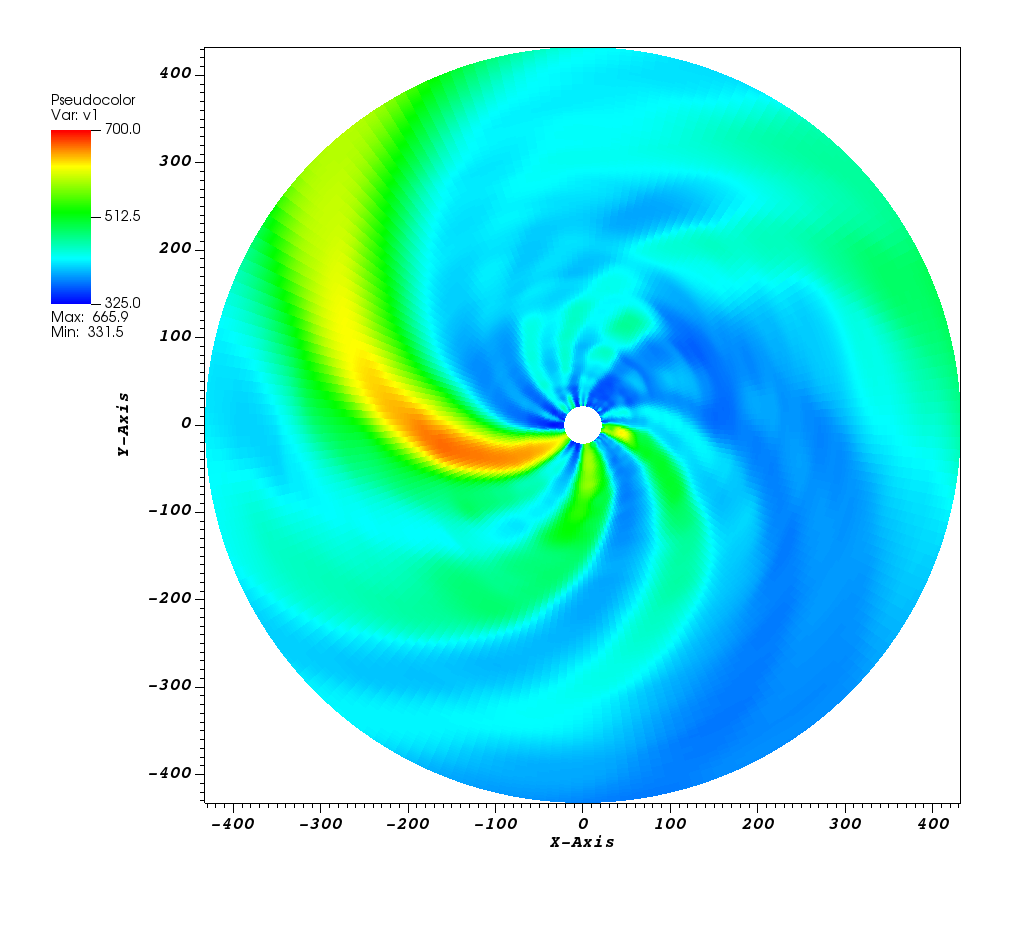}

     \end{subfigure}
     \hfill
     \begin{subfigure}[b]{0.33\textwidth}
         \centering
         \includegraphics[width=\textwidth]{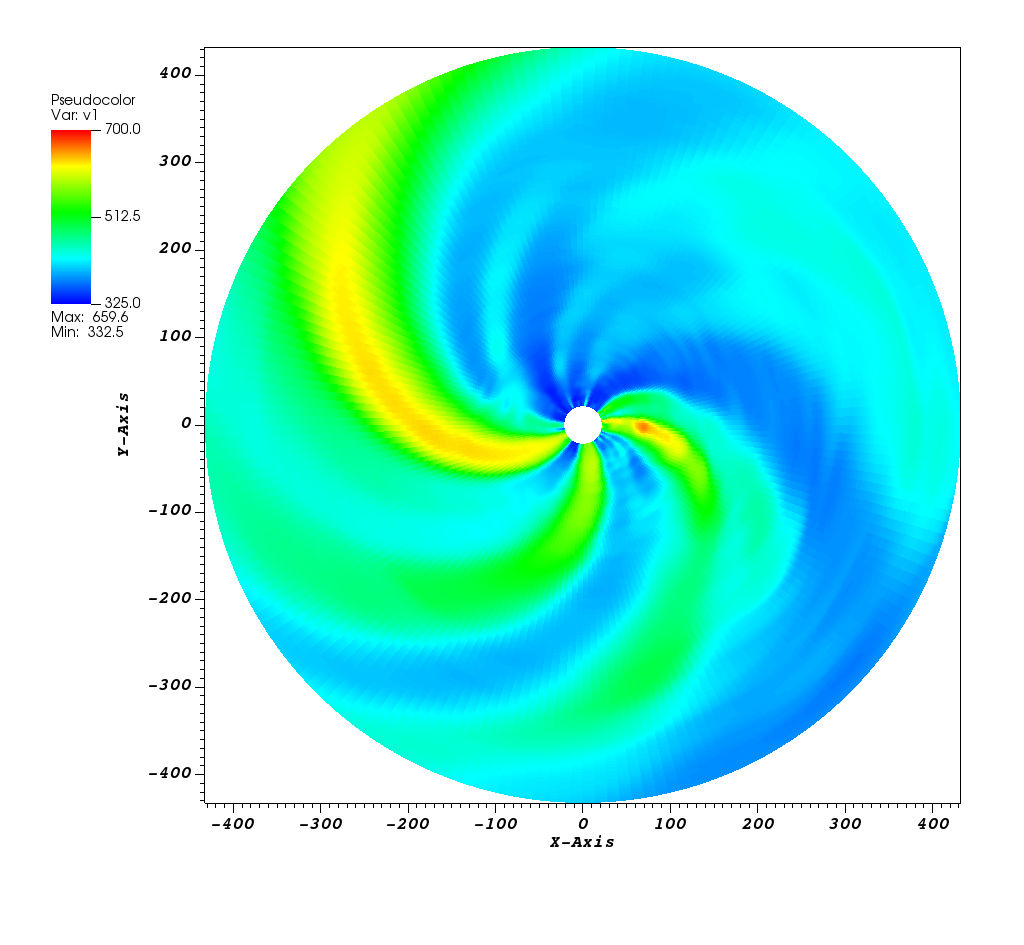}

     \end{subfigure}
     \hfill
     \begin{subfigure}[b]{0.33\textwidth}
         \centering
         \includegraphics[width=\textwidth]{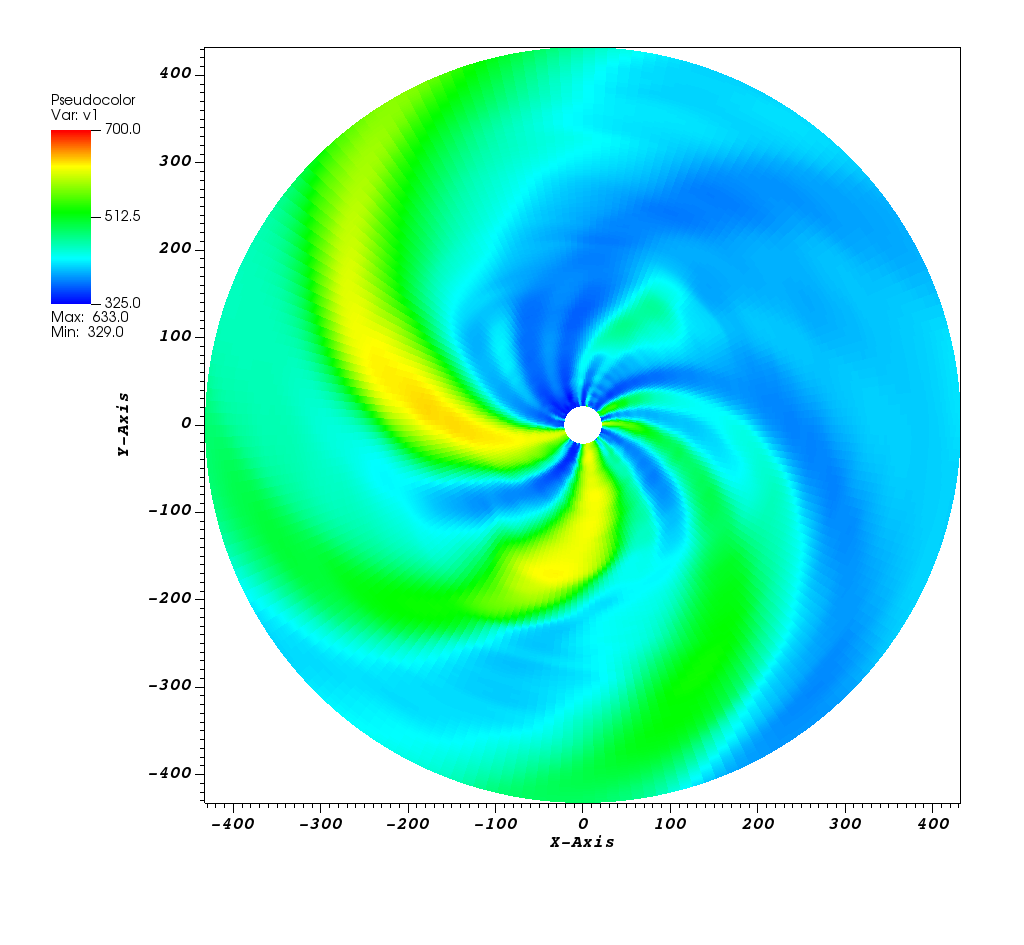}

     \end{subfigure}
     \hfill
     \begin{subfigure}[b]{0.33\textwidth}
         \centering
         \includegraphics[width=\textwidth]{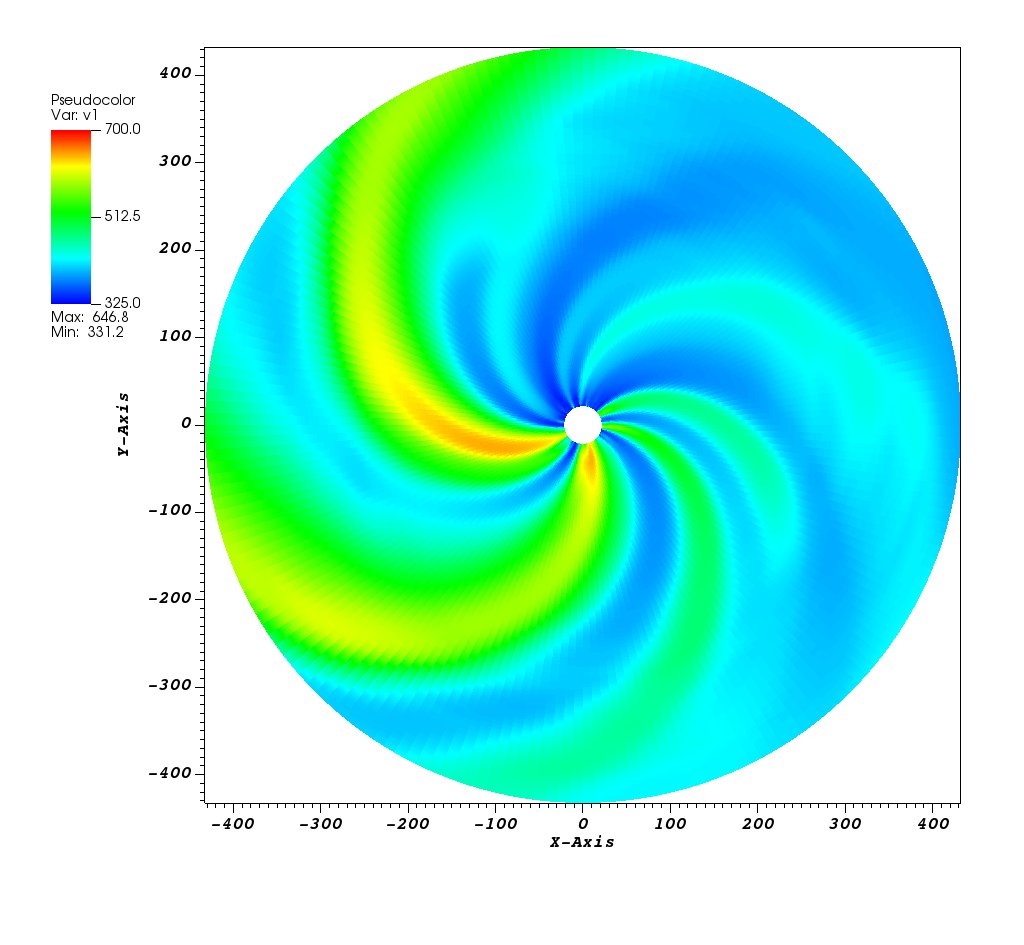}

     \end{subfigure}
  \caption{The equatorial planes from the same Icarus simulation at 15, 20, 25, 30, 35, and 40 days in the simulation from the upper left panel to the bottom right panel, respectively. The radial velocity values are plotted with the corresponding colour map values in [km s$^{-1}$]. }\label{fig:dynamic_wind_2019}
\end{figure*}

The boundary condition treatment was generalized when upgrading to \texttt{Icarus 3.0}. The original description of the boundary conditions is given in \cite{Verbeke2022}. \texttt{Icarus} reads the standard coronal boundary file generated by the WSA model, which the EUHFORIA heliospheric model also uses. The inner radial boundary condition treatment update was hardcoded in the user file, and it was assumed that the number of ghost cells would be fixed to two in the previous version. This, however, precluded the use of many flavours of higher order limiters and schemes available in \texttt{MPI-AMRVAC} that require a larger stencil.

\cite{Keppens2023} advocated using temporally and spatially varying boundary data stored in a lookup table for updating driven boundary conditions. The needed input for \texttt{Icarus~3.0} as boundary data that evolves at $r=0.1$ AU was generated in Visualisation ToolKit (VTK) format, as demonstrated in the example of \cite{Keppens2023}. To achieve this, a Python script was developed to convert the standard output of the WSA model into the ordered VTK file. Since, in this first step, we are upgrading the method of reading stationary boundary conditions, 2D data corresponding to the 0.1~AU shell were written to the input boundary file. 

The initial conditions describing the solar wind in \texttt{Icarus} are also interpolated on the computational grid from the provided boundary file. The first snapshot present in the boundary file is taken for the initialisation. The values for radial velocity, number density, pressure, and radial magnetic field are set at the inner boundary, and simple radial expansion laws are obeyed to extrapolate them to set initial values throughout the entire domain, the same way as before. In the standard output of the WSA model, the temperature is stored in the boundary file. Updating the boundary conditions through lookup tables uses primitive variables, thus the temperature is converted to pressure in the pre-processing when creating the inner boundary file in the VTK format. In the standard boundary treatment with the input boundary data, only the data read from the boundary file can be imposed in the simulation. Therefore, only the solar wind is modelled using the updated approach without injecting a CME as a first step. The inner boundary is now updated with the {\fontfamily{pcr}\selectfont bc\_data} setting from the parameter file, which requires access to the inner boundary conditions. Since the WSA model only provides radial velocity, number density, temperature, and radial magnetic field, the inner boundary conditions for other quantities are set as default to the ``{\fontfamily{pcr}\selectfont asymm}'' boundary condition that enforces them to vanish at the boundary surface corresponding to 0.1~AU. The parameter list in the input file includes the file name containing information about the inner boundary, the setting of the type of interpolation, and whether the variables are primitive or conservative.

This way, the inner boundary conditions are set from the lookup table created based on the input boundary file. Figure~\ref{fig:amrvac3_old_new} shows a comparison between the old (teal) and the new (orange) ways of implementing the boundary conditions in \texttt{Icarus 3.0}. Both simulations are performed on a low-resolution uniform grid. The differences between the two curves are minimal, as the interpolation in the two cases is slightly different. In the old implementation, the values are linearly interpolated in both ghost cells, while in the new one, the same values are set.

\section{Dynamic Heliosphere driving}
\label{dynamic_bcs}

So far, we presented results that, in essence, reproduce and slightly improve on previous \texttt{Icarus} versions by using the upgrades provided by the underlying framework, but that still handles a steady solar wind background.
Steady solar wind driving limits the accuracy of the relaxed solution in the 3D MHD heliosphere. As a main novelty to the \texttt{Icarus} model, we now realise inner boundary driving updated from steady to dynamic to enable long-term solar wind forecasting. In the newly implemented setting, the 3D MHD heliospheric model is driven with the input generated by the WSA model, but now with evolving magnetograms. As the WSA-like coronal model available in the EUHFORIA package only calculates solar wind conditions for a single magnetogram, the coronal model was automated to retrieve the magnetograms between the dates of interest and to calculate the plasma conditions at 0.1~AU. The output files of the WSA model are stacked as a time series in the inner boundary file. Instead of the previously generated 2D solar wind boundary file, a 3D file is generated, where the third dimension accounts for the time, and the 2D boundary data are stored for each magnetogram timestamp. The timestep in the simulation is naturally smaller than the magnetogram cadence. Therefore, the values are linearly interpolated between the consecutive snapshots. The step-by-step procedure for obtaining the time-dependent solar wind conditions in \texttt{Icarus} is the following:
\begin{enumerate}
    \item Determine the start of the period of interest.
    \item Identify the period's start and end points for the automated WSA model:
    \subitem t$_{start}$ = t$_{interest}$ - relaxation time,
    \subitem t$_f{end}$ = t$_{interest}$ + time period.
    \item Run the WSA model with a cadence of choice (minimum 1~h).
    \item Generate an input 3D boundary file for \texttt{Icarus} in the VTK format.
    \item Run with {\fontfamily{pcr}\selectfont bc\_data} boundary conditions.
\end{enumerate}

The relaxation time in the simulations presented below was fixed to 8 days but can be modified by the user. The relaxation time is set the same way in EUHFORIA and corresponds to the time that slow solar wind requires to reach the outer boundary of the inner heliosphere. In the current version, the input VTK file is generated with all the information before the start of the simulation and the boundary is updated from this file. In what follows, we showcase how this dynamic solar wind evolution can dramatically increase the forecasting abilities measured by direct comparisons with in-situ data.

\subsection{Solar wind: September-October 2019}
\label{2019case}
\begin{figure}[hpt!]
    \includegraphics[width=0.5\textwidth]{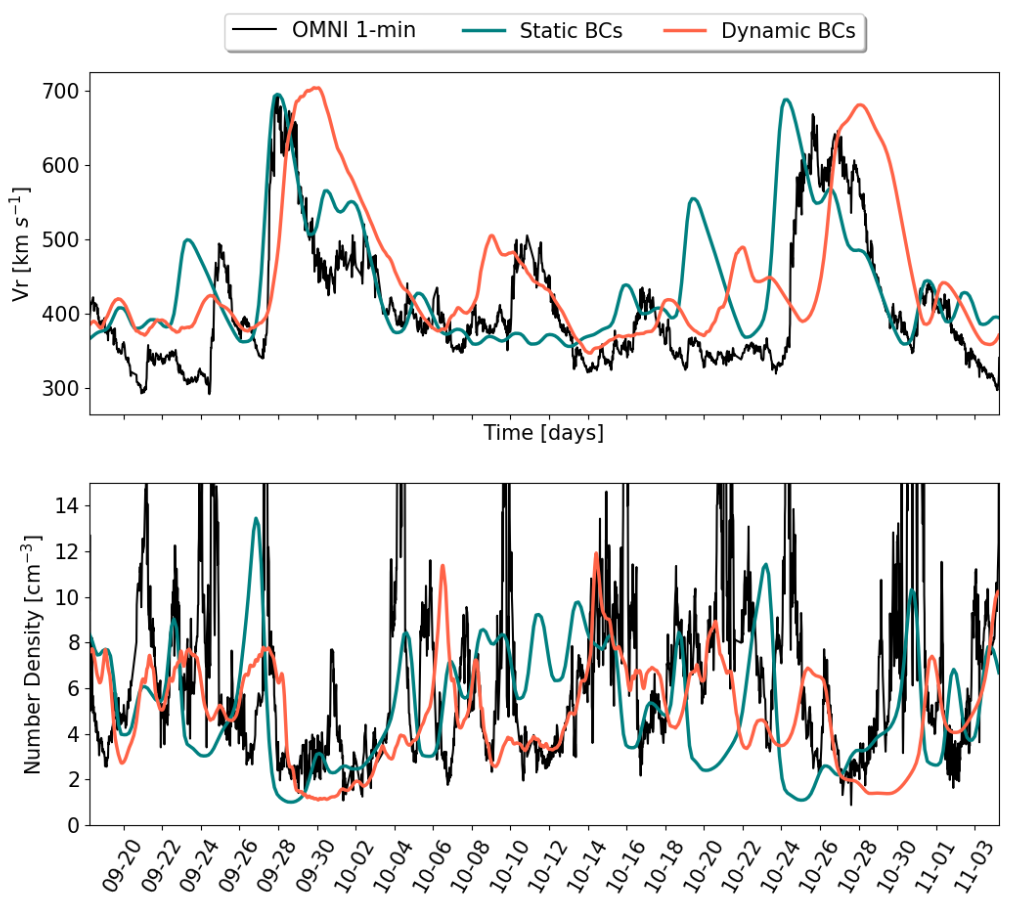}
  \caption{Radial velocity (upper panel) and number density (lower panel) at Earth simulated with \texttt{Icarus 3.0} with static (teal) and dynamic (orange) inner boundary driving. The simulations are performed on a medium-resolution computational domain defined in \cite{Baratashvili2022}. Observed data are plotted in black. The data is from September and October 2019. }\label{fig:earth_static_dynamic_2019}
\end{figure}

\begin{figure}[hbt!]
    \includegraphics[width=0.5\textwidth]{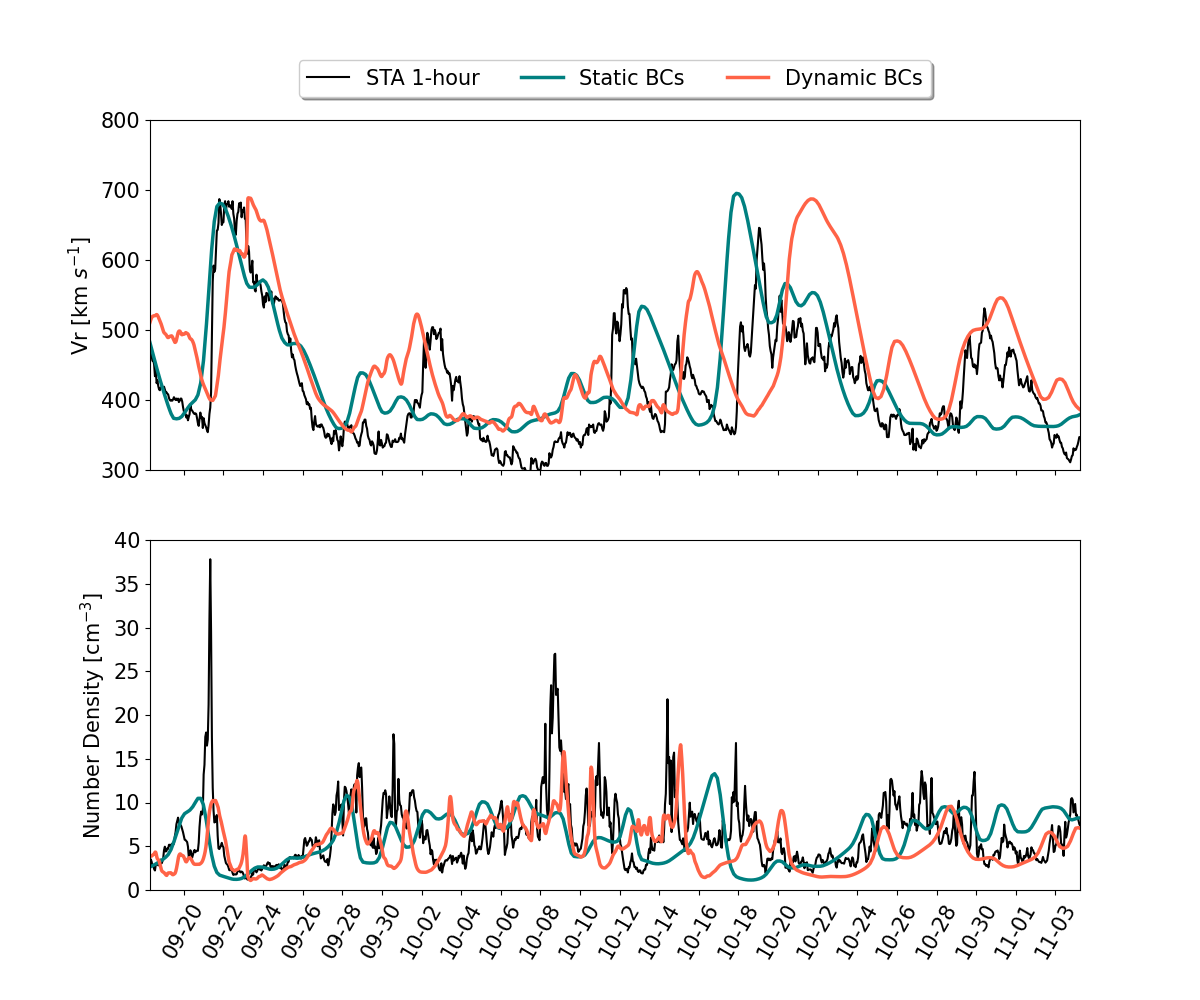}
  \caption{Radial velocity (upper panel) and number density (lower panel) at Stereo A performed with \texttt{Icarus 3.0} with static (teal) and dynamic (orange) inner boundary driving. The simulations are performed on a medium-resolution computational domain defined in \cite{Baratashvili2022}. Observed data are plotted in black.  The data is from September and October 2019.} \label{fig:sta_static_dynamic_2019}
\end{figure}

The results of the dynamic heliospheric driving are presented with the solar wind modelling for the September-October 2019 case. \citet{Wijsen2021} studied this period, which is an interesting case since one rather sizeable coronal hole was observed on the solar surface.
If we want to model the solar wind and compare to observations from t$_{interest}$ = 2019.09.18 06:14:00, then from the standard procedure presented in the previous section we can calculate the start and the end of the simulation time: t$_{start}$ = 2019.09.10 06:14:00 and t$_{end}$ = 2019.10.18 06:14:00. Eight days before is chosen to allow the wind to travel to the outer boundary, which is similar to the previous ``relaxation'' phase in the steady simulations. However, the wind obtained after the ``relaxation'' phase is not steady. For the first test, we chose the cadence to be two hours, as 2019 was close to the solar minimum, and we modelled one month. These settings produce 454 coronal boundary files, later stacked in the single input boundary VTK file. The simulation was done with the generated boundary file containing the updated magnetograms. All simulations are performed on a medium-resolution uniform computational domain defined in \cite{Baratashvili2022} with 600 cells in the radial direction and two-degree resolution in the latitudinal (64 cells) and longitudinal (192 cells) directions. The simulation time was set to 42 days. Figure~\ref{fig:dynamic_wind_2019} shows the solar wind radial speed profiles in the equatorial plane from the time-dependent simulation with updating magnetograms. The panels in the first row show the snapshots 15, 20, and 25 days after the start of the simulation; the second row corresponds to the solar wind configuration after 30, 35, and 40 days in the simulation. The updated solar wind instead of the steady, relaxed solution can be seen. Figures~\ref{fig:earth_static_dynamic_2019} and \ref{fig:sta_static_dynamic_2019} show the results for the simulations performed with a fixed magnetogram (teal) and updating magnetograms (orange) at Earth and Stereo A, respectively. The observed data are plotted in black. The results are compared visually instead of running the various models due to noise in the observed data. In Figure~\ref{fig:earth_static_dynamic_2019}, we can see that the static case better captures the arrival time for the high-speed stream on $\sim28$ September.
Moreover, as mentioned earlier, the coronal model was modified to account for this high-speed stream in \cite{Wijsen2021}, and the same modified WSA parameters are used to generate all boundary files with the WSA model. The high-speed stream is missed in the steady simulation starting on October 10, whereas, in the time-dependent one, we see an increase in the speed profile and a decrease in the number density data. The slow wind path between 5 and 7 October is captured in the time-dependent simulation, and a sharp increase in density is also present. We can see that the number density profiles are more similar to the observed data in the time-dependent simulation than in the steady one.

Figure~\ref{fig:sta_static_dynamic_2019} shows the results from the same simulations but at Stereo A. In the upper panel, we can see that the arrival of the first high-speed stream is again better modelled by the static inner boundary condition simulation. However, the profile of the dynamic one shows a step increase in the radial velocity values, which is also present in the observed data. The static boundary simulation misses the second high-speed stream starting from October 1. The velocity profile is more similar to the observed data in the dynamic simulation, but the peak is achieved earlier. The three higher-speed streams on October 12, 14 and 18 are modelled better by the time-dependent simulation than the static simulation, as all three peaks are present in the time series. However, the agreement with the observed data is not good. Starting on October 30, the last high-speed stream is modelled accurately by the dynamic simulation, while we do not see any signatures of higher-speed regions in the steady-simulation results. {As we can see from the results in figures~\ref{fig:earth_static_dynamic_2019} and~\ref{fig:sta_static_dynamic_2019}, the solar wind is not automatically improved everywhere compared to the observed data when switching to the dynamic boundary conditions from the steady boundary driving. However, we can see a significant improvement. Therefore, upgrading heliospheric modelling from steady regime to dynamic is a step toward more realistic and continuous solar wind modelling. }

Table~\ref{table:run_times_2019} shows the time simulations required when performed on six nodes with two Xeon Gold 6240 CPUs@2.6 GHz (Cascadelake), 18 cores each, on the Genius cluster at KU Leuven. The simulations with static boundary driving took 59 minutes, and those with dynamic boundary driving took 57 minutes. The difference between execution times is not significant. 

\begin{table}[htb!]
  \caption{Run times (wall-clock time) for \texttt{Icarus 3.0} simulations.}
  \centering
   \begin{tabular}{c c  }
  \hline\hline
   Simulation & Time \\[4pt]
   \hline
Steady & 0 h 59 m \\[4pt] 
\hline
Dynamic & 0 h 57 m \\[4pt] 
\hline
 \end{tabular}
 \tablefoot{ All the simulations were performed on six nodes with two Xeon Gold 6240 CPUs@2.6 GHz (Cascadelake), 18 cores each, on the Genius cluster at KU Leuven.}
  \label{table:run_times_2019}

\end{table}

\subsection{Solar wind and CMEs: June-July 2015}
\label{2015case}
\begin{table*}[t!]
\caption{CME model parameters of the five simulated eruptions.}   
\label{table:2015_cme_parameters}   
\centering            
\begin{tabular}{c c c c c c}         
\hline
CME \# & Time at 0.1~AU & Latitude  & Longitude  & Half-width & Speed   \\ 
& & deg (HEEQ) & deg (HEEQ) & deg & km s$^{-1}$ \\
\hline\hline
  1 & 2015-06-18T20:00:00 &11 &-50 &45 &1000.0  \\
  2 & 2015-06-19T14:59:00 & -33& 9 & 54& 603.0 \\
  3 & 2015-06-21T05:01:00 &7 & -8& 47& 1250.0 \\
  4 & 2015-06-22T21:10:00 &14 &3 & 45& 1155.0 \\
  5 & 2015-06-25T10:51:00 & 23& 46& 41& 1450.0 \\
\hline                                  
\end{tabular}
 \tablefoot{ The parameters are taken from the DONKI catalogue and used by \cite{Pomoell2018}.}
\end{table*}
\begin{figure}[hpt!]
    \includegraphics[width=0.5\textwidth]{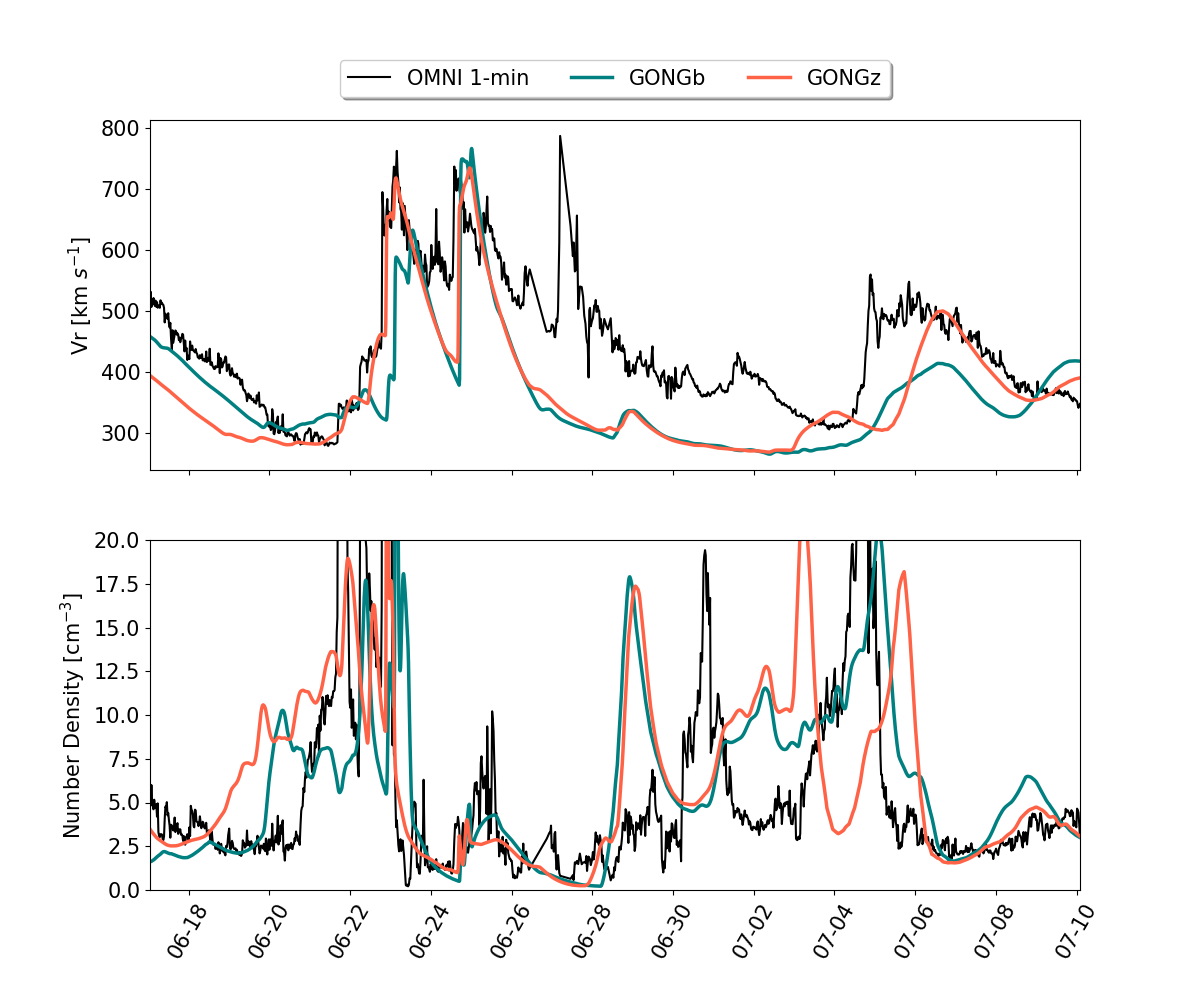}
    \includegraphics[width=0.5\textwidth]{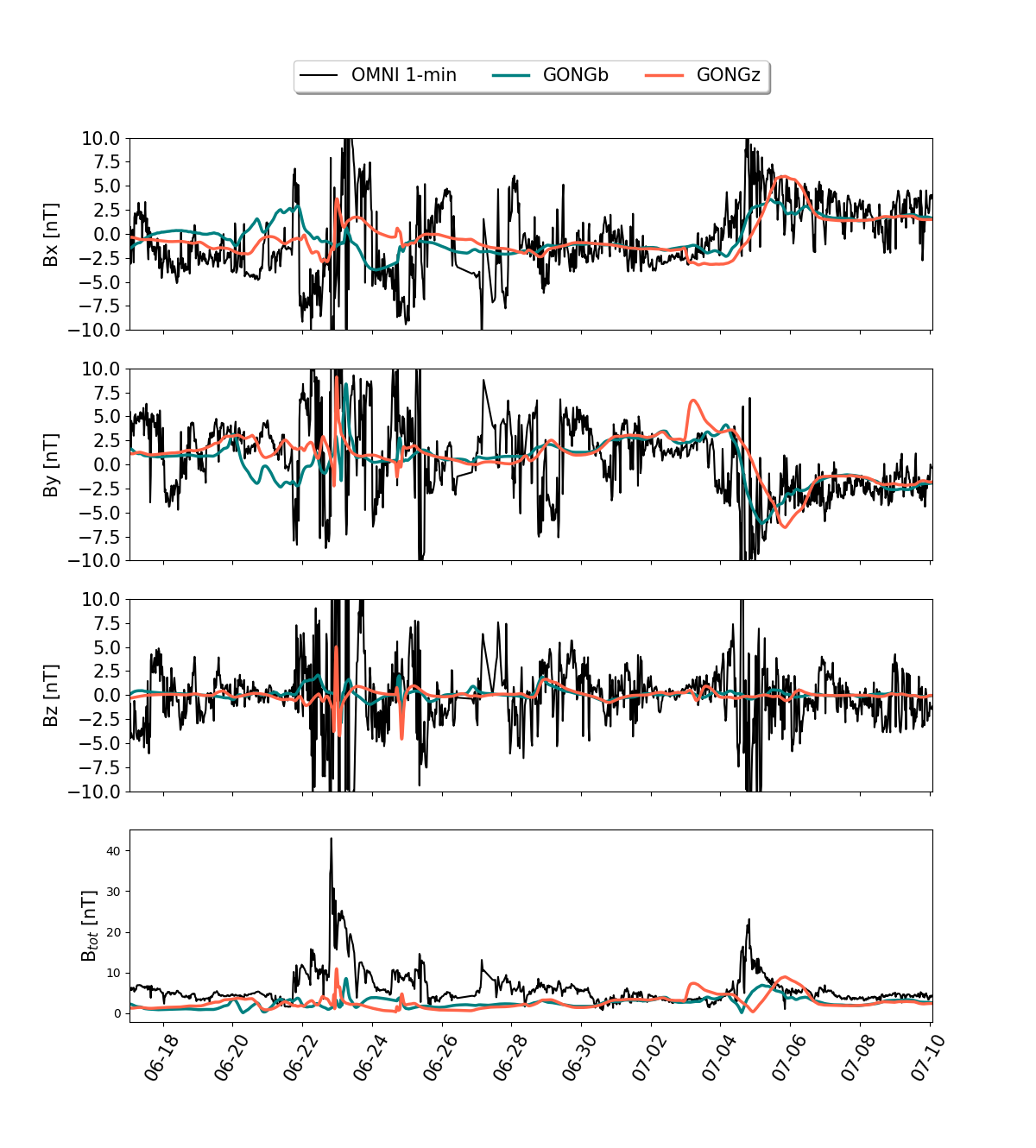}
  \caption{Medium resolution simulation results performed with \texttt{Icarus 3.0} with GONGb (teal) and GONGz (orange) products. Upper panel: Radial velocity (upper) and number density (lower). Lower panel: $B_x$, $B_y$, $B_z$ and total magnetic field from top to bottom. Observed data are plotted in black and correspond to the June-July period 2015. }\label{fig:earth_gongb_gongz}
\end{figure}
To validate the dynamic solar wind model more extensively, we consider a second case corresponding to June-July 2015. This case was considered in \cite{Pomoell2018}, where they also injected five consecutive CMEs on top of the modelled solar wind in EUHFORIA. The WSA model is computed for the static boundary conditions used in the paper corresponding to the GONG magnetogram on 25.06.2015 01:04 UT. 
\begin{figure*}[hbt!]
\centering
    \begin{subfigure}[b]{0.33\textwidth}
         \centering
         \includegraphics[width=\textwidth]{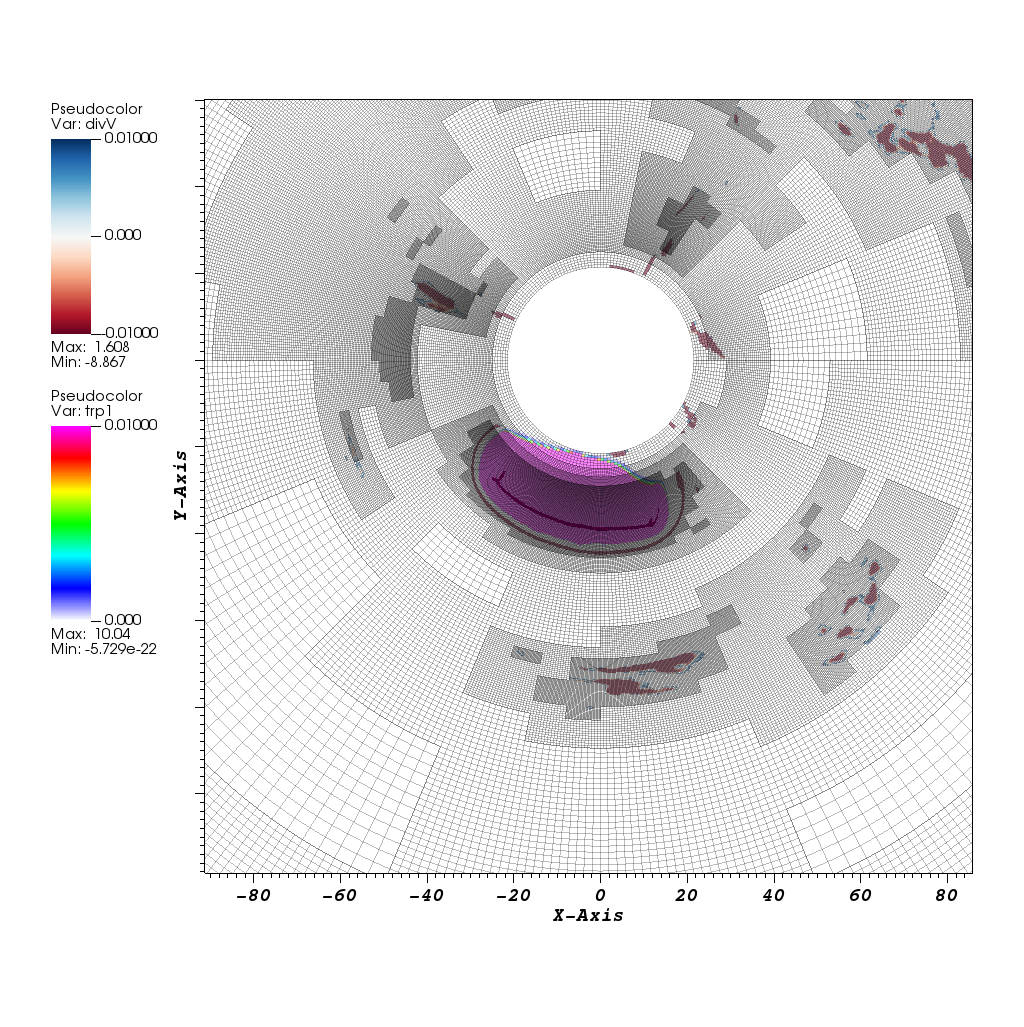}

     \end{subfigure}
     \hfill
          \begin{subfigure}[b]{0.33\textwidth}
         \centering
         \includegraphics[width=\textwidth]{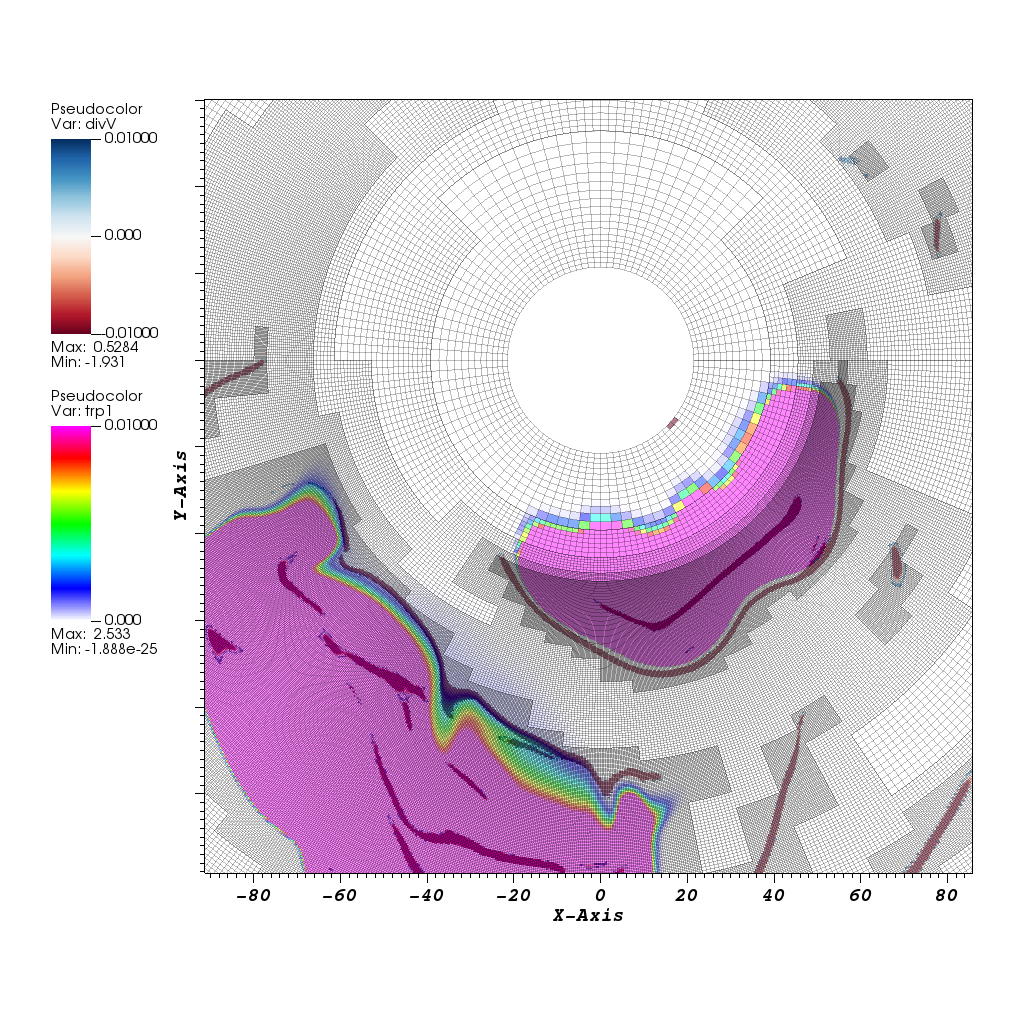}

     \end{subfigure}
     \hfill
          \begin{subfigure}[b]{0.33\textwidth}
         \centering
         \includegraphics[width=\textwidth]{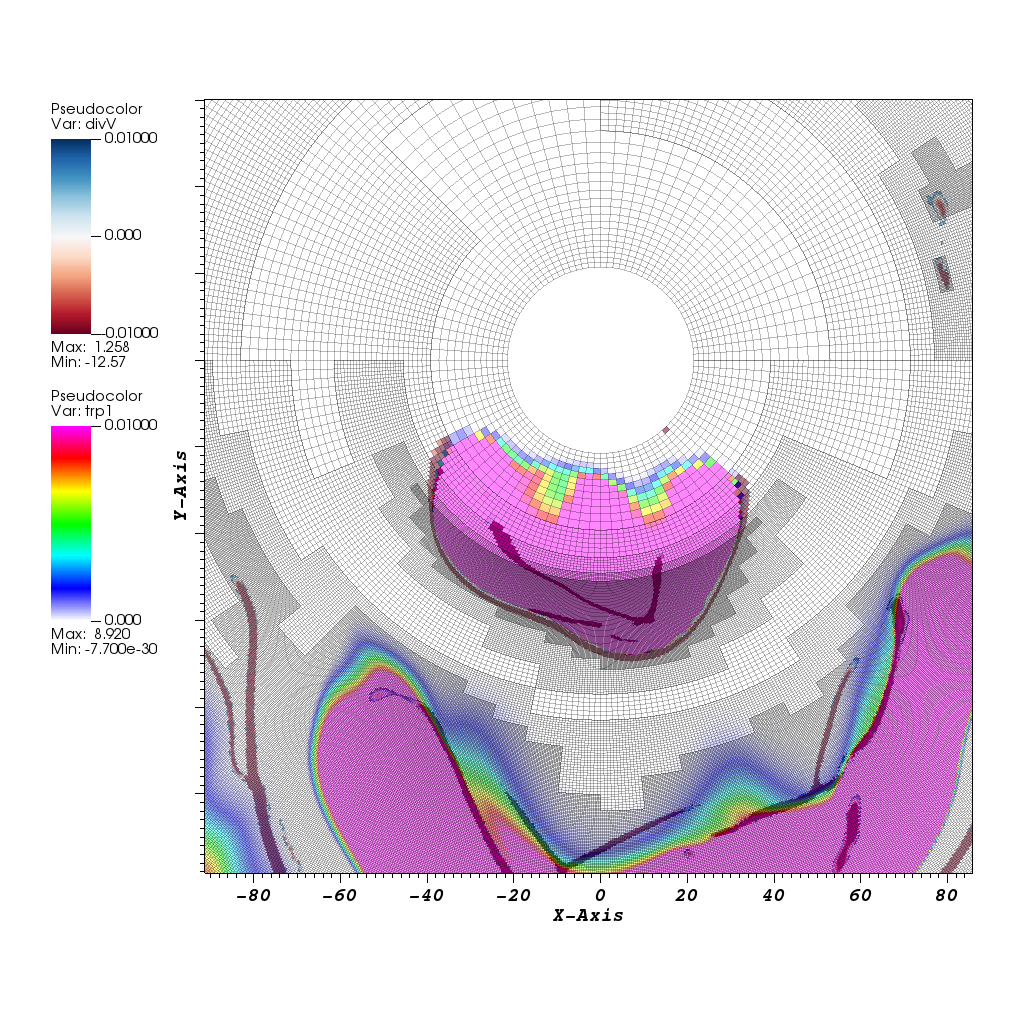}

     \end{subfigure}
     \hfill
     \begin{subfigure}[b]{0.33\textwidth}
         \centering
         \includegraphics[width=\textwidth]{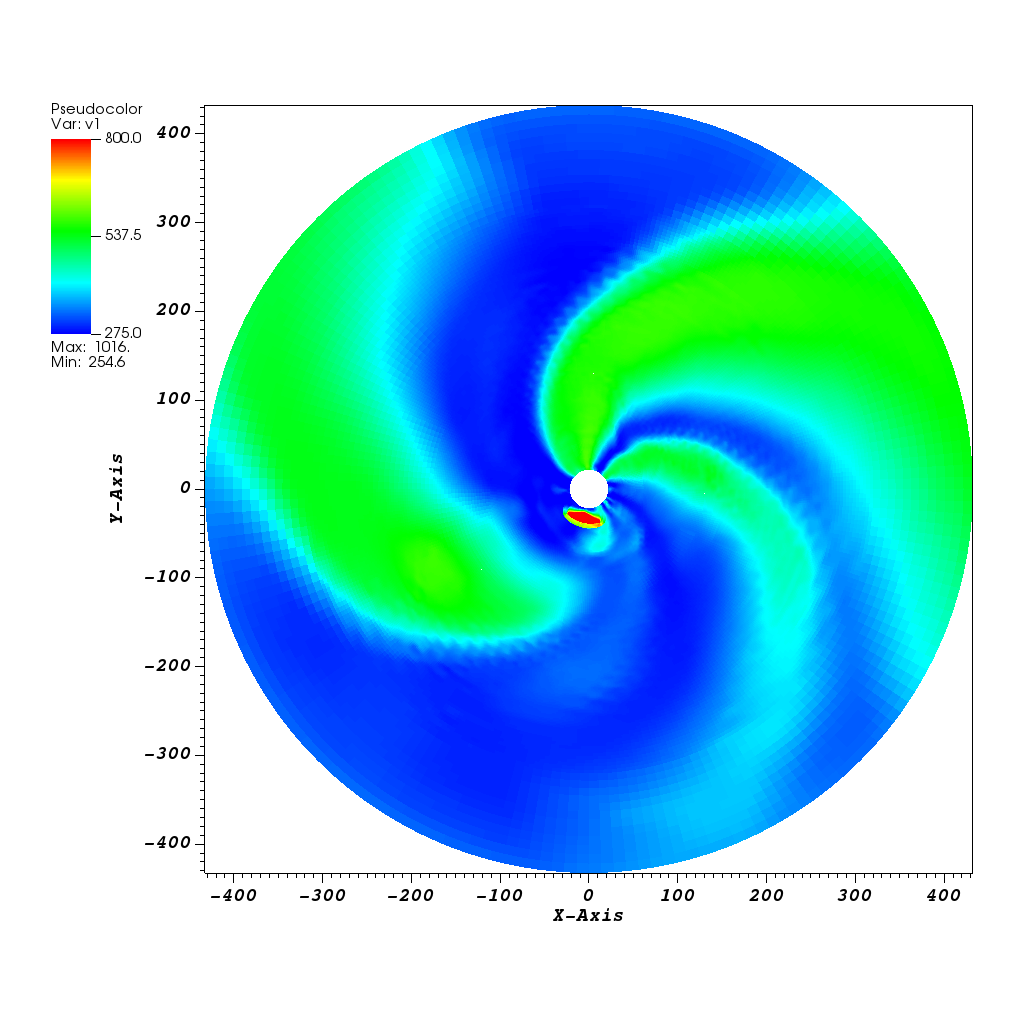}

     \end{subfigure}
     \hfill
     \begin{subfigure}[b]{0.33\textwidth}
         \centering
         \includegraphics[width=\textwidth]{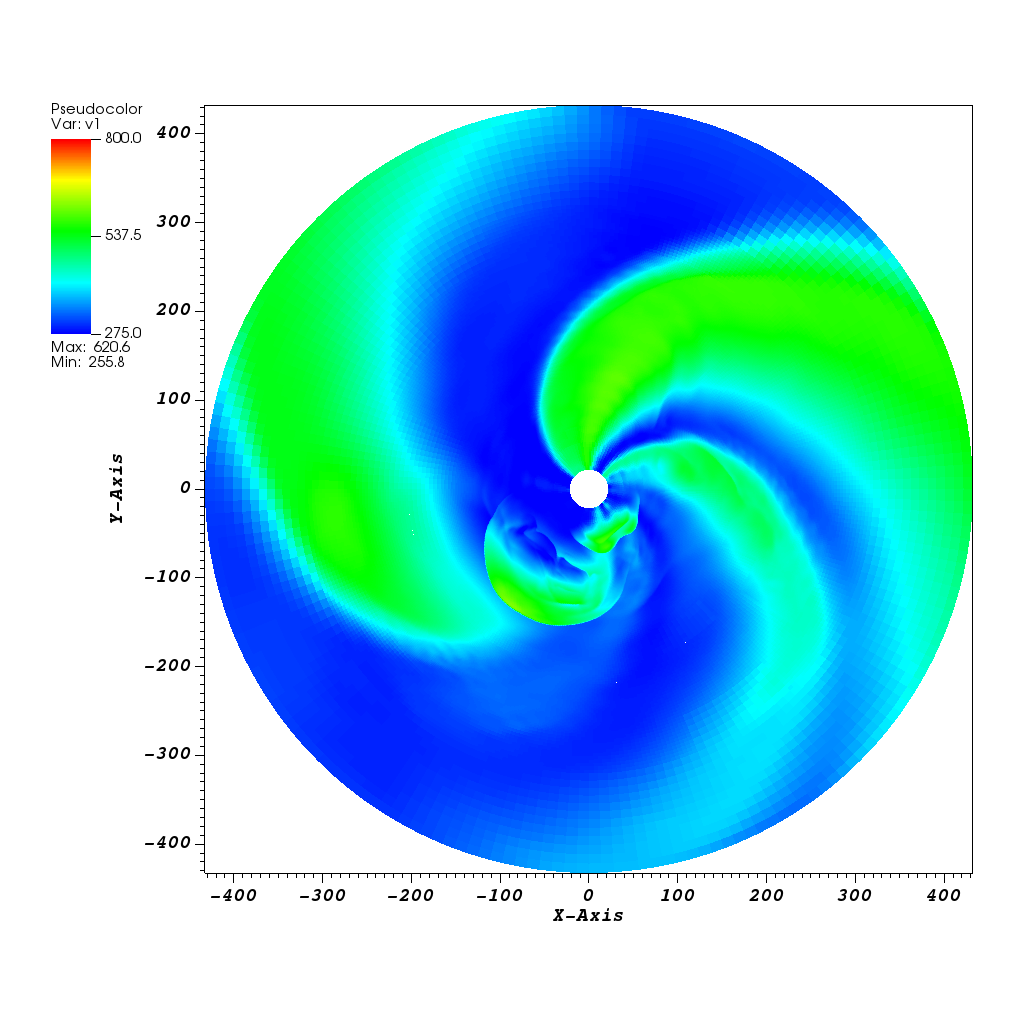}

     \end{subfigure}
     \hfill
     \begin{subfigure}[b]{0.33\textwidth}
         \centering
         \includegraphics[width=\textwidth]{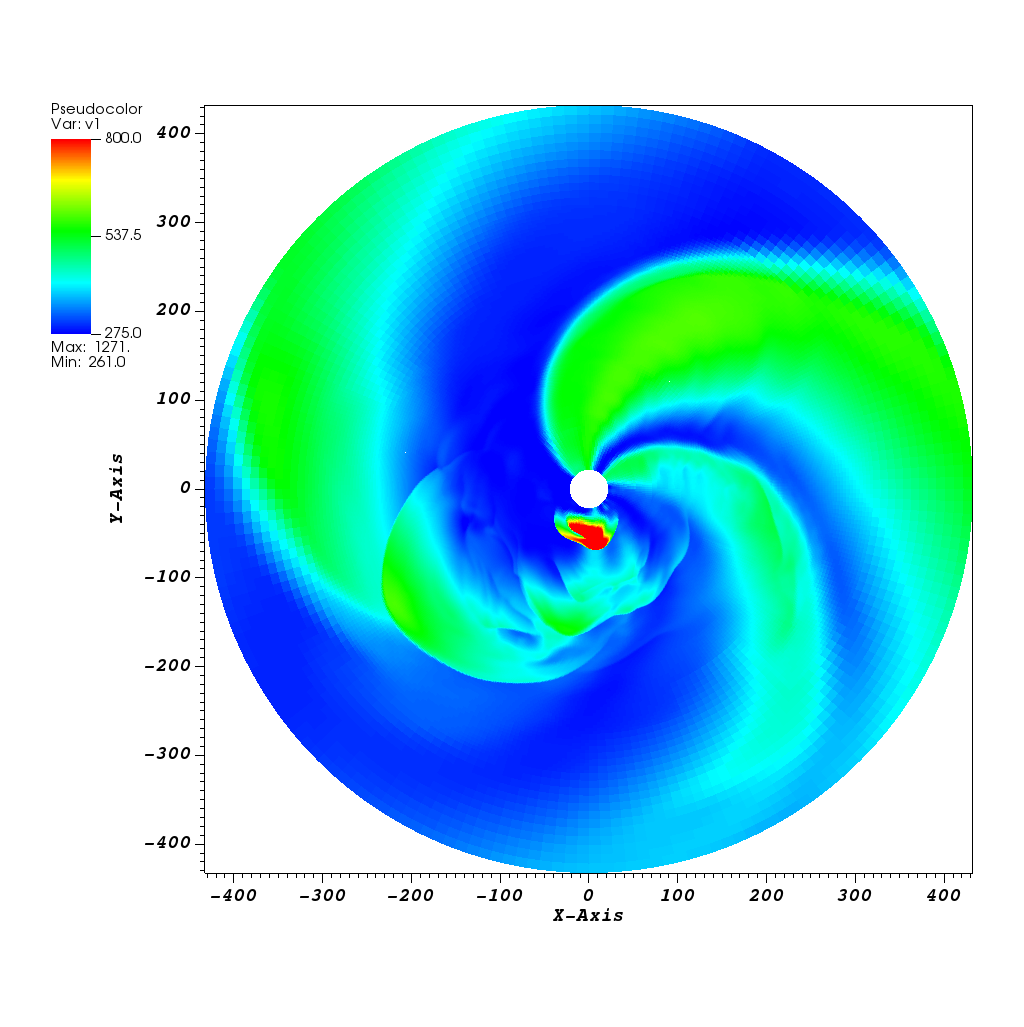}

     \end{subfigure}
     
  \caption{The equatorial planes from the same \texttt{Icarus} simulations at 9, 10.3, and 11.5 days in the simulation. The upper row shows the CME density and  $\nabla\cdot\mathbf{V}$ values in the code units overplotted with the refined mesh. The bottom row shows the radial velocities for the same snapshots. }\label{fig:dynamic_wind_2015}
\end{figure*}

As mentioned before, driving the boundaries with the standard {\fontfamily{pcr}\selectfont bc\_data} setting from the parameter file only reads the data stored in the boundary file, and the inner boundary is not modified further in the simulation. This way, the data cube can be passed to \texttt{Icarus}, but the CME can not be superposed from the inner boundary during the simulation. Since modelling CMEs is crucial for space weather forecasting, we generalized the standard {\fontfamily{pcr}\selectfont bc\_data} setting to {\fontfamily{pcr}\selectfont bc\_icarus}. The {\fontfamily{pcr}\selectfont bc\_icarus} setting was implemented to combine the {\fontfamily{pcr}\selectfont bc\_data} and manual adjusting of the inner boundary to inject the CME in the simulation domain. A special {\fontfamily{pcr}\selectfont icarus\_list} segment is added to the \texttt{MPI-AMRVAC~3.0} parameter settings to deal with the boundary driving in \texttt{Icarus}. The settings include the adaptive mesh refinement (AMR) criterion to be applied to the computational domain, the CME activation flag, the number of CMEs, the relaxation and CME insertion periods for the solar wind, the CME parameter file name and the magnetogram timestamp.
This list considers all specific parameters necessary for setting up simulations with or without a CME injection. The default refinement criteria are implemented following \cite{Baratashvili2022}, namely ``{\fontfamily{pcr}\selectfont tracing}'', ``{\fontfamily{pcr}\selectfont shock}'', and ``{\fontfamily{pcr}\selectfont combined}''.We recall that these options were found to work well for tracing the entire CME cloud or, focusing on the compression regions around shocks, or both. The passed ``{\fontfamily{pcr}\selectfont cme flag}'' setting indicates whether the CMEs will be injected from the inner heliospheric boundary, relaxation and CME insertion options are defined the in the same way as \cite{Pomoell2018} and are given in days.

When the CME flag is deactivated, {\fontfamily{pcr}\selectfont bc\_icarus} updates the inner boundary conditions the same way as {\fontfamily{pcr}\selectfont bc\_data}. If the CME flag is activated, then special boundary conditions are activated, which check where the CME must be injected and modify only these points at the inner boundary from the parameters in the indicated CME parameter file. In this way, the CME can be injected into the domain. 

The selected solar wind period was modelled with five CMEs to test and validate the newly implemented boundary conditions. Table~\ref{table:2015_cme_parameters} shows the parameters for introducing the CMEs at the inner boundary. The start and the end times of the simulation are calculated from the magnetogram timestamp and correspond to t$_{start}$ = 2015.06.10 01:14:00 and t$_{end}$ = 2015.07.09 22:14:00 since the relaxation time corresponds to 8 days and the CME insertion to 7 days, similar to \cite{Pomoell2018}.

For computing the WSA coronal model, the pipeline automatically retrieves the Standard QuickReduce Magnetogram Synoptic Maps (denoted by GONGb). During the solar minimum between cycles 23-24 \citep{Jian2016}, a Zero-point Corrected QuickReduce Synoptic Map Data (denoted with GONGz) was developed, which improves the quality of the magnetograms by adjusting the zero-point uncertainties that are caused by the magnetogram modulators \citep{Hill2018}. \cite{Li2021} considers the effects of the different products from GONG magnetograms. 

Figure~\ref{fig:earth_gongb_gongz} compares the results for the heliospheric simulations using the standard (teal) or zero-corrected (orange) magnetogram for the coronal model. The simulations are performed with the new dynamic boundary driving. The observed data are plotted in black. The upper panel compares the modelled radial velocity and number density values, whereas the lower panel focuses on the magnetic field configuration. The CME model used in this case is not magnetized, therefore the magnetic field is underestimated when the CMEs pass through Earth. However, when looking at the regions before and after the CME signatures, the modelled data with the zero-point corrected magnetogram better agrees with the observed data than the standard case. In particular, the region between 05.07-07.07 is modelled better with the GONGz product than the GONGb product. Therefore, to assess the results of our simulations, we chose the GONGz product. 

\begin{figure*}[hpt!]
    \includegraphics[width=0.5\textwidth]{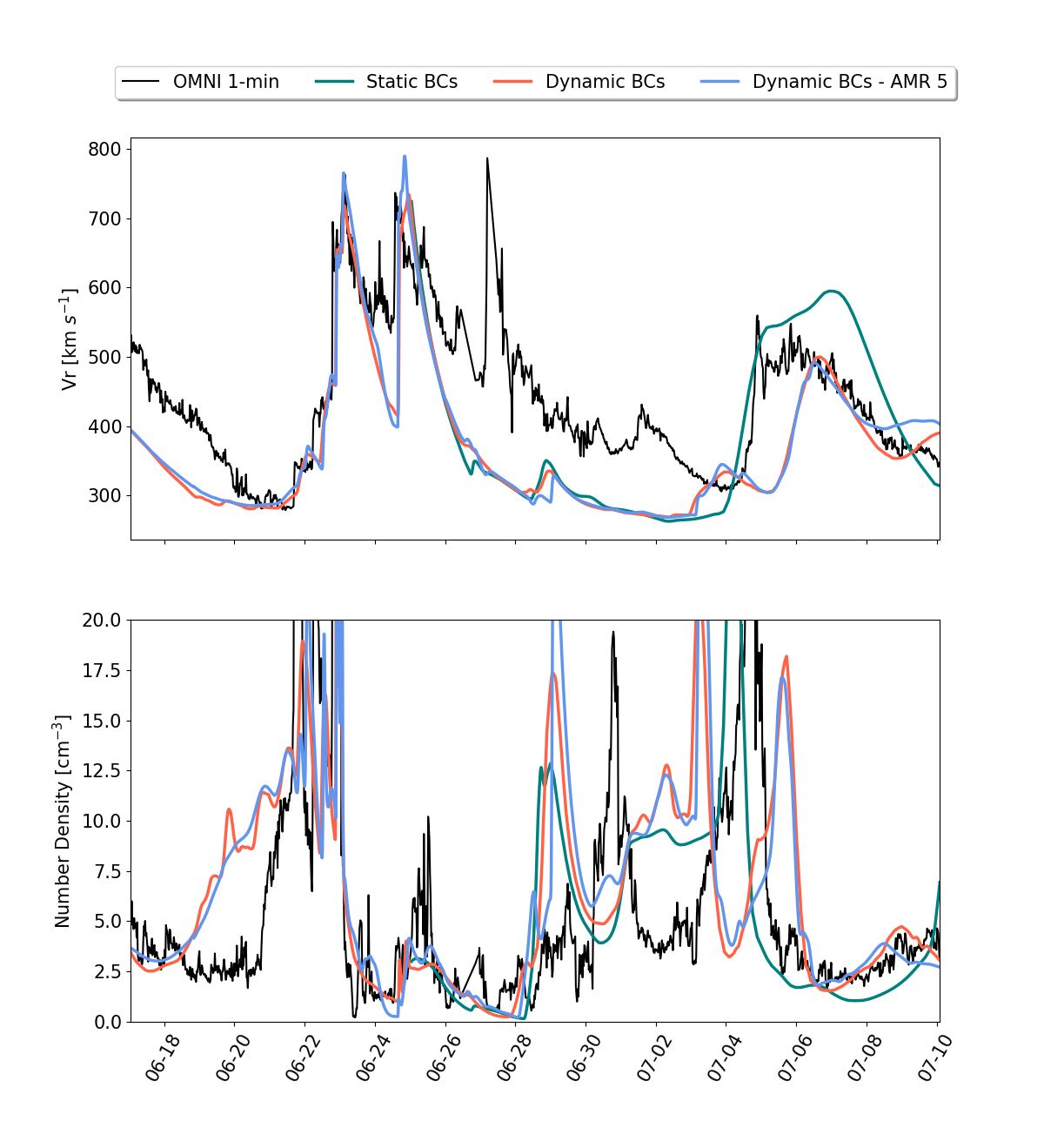}
    \includegraphics[width=0.5\textwidth]{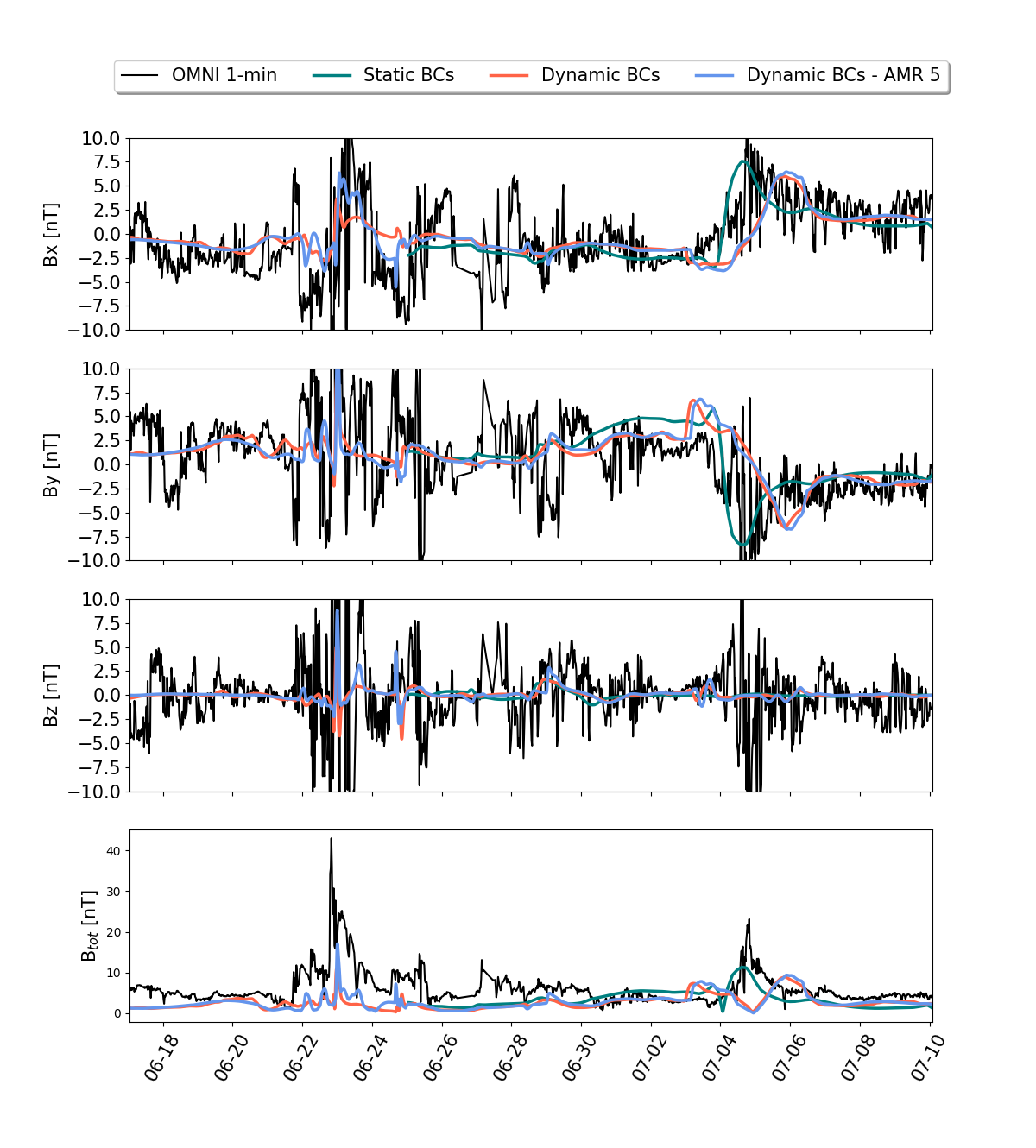}
  \caption{Medium resolution simulation results performed with \texttt{Icarus 3.0} with the steady and dynamic boundary driving. Left panel: Radial velocity (upper) and number density (lower). Right panel: $B_x$, $B_y$, $B_z$ and total magnetic field from top to bottom. Observed data are plotted in black and correspond to the June-July period 2015. }\label{fig:earth_2015_steady_dynamic}
\end{figure*}

Figure~\ref{fig:dynamic_wind_2015} shows the modelled heliosphere with the newly developed  {\fontfamily{pcr}\selectfont bc\_icarus} boundary conditions. AMR level 5 is applied with the ``{\fontfamily{pcr}\selectfont combined}'' refinement criterion. The results of the simulation are plotted in the equatorial plane. The upper row shows the divergence of velocity and CME density values with the upper and lower colour maps on the left of the figures, respectively. The values are overlaid with the computational grid with the AMR level 5 grid fixed from the parameter file. The snapshot is zoomed closer to the inner heliospheric boundary to better show the different AMR levels in the domain. This AMR criterion refines the computational grid to a higher resolution if the CME or the shock is present in the given block of the domain. Therefore, we can see that other areas apart from the CME density enhancements are also refined to higher AMR levels. The bottom row shows the radial velocity values in the equatorial plane. The snapshots are zoomed out to show the whole domain from the inner to the outer boundary. The zoomed-in CMEs in the upper row can be well identified with the enhanced velocity values in the domain in the bottom row.

Figure~\ref{fig:earth_2015_steady_dynamic} shows the time series at Earth. The black curve corresponds to OMNI data. The simulation results from steady boundary driving are given in teal colour, the dynamic boundary driving results are plotted in orange, and the dynamic simulation with five levels of AMR applied to it is plotted in blue. The left figure shows the radial velocity and number density values, while the right panel shows the magnetic field components and the total magnetic field. As there are five CMEs injected in the simulation, the times when the CMEs interact with the Earth cannot be considered when assessing solar wind modelling as the impact of the CME is stronger. We can notice that the difference is small at the beginning of the time series between 
the dynamic simulation on a uniform grid and with five levels of AMR. Before the CMEs arrive, the AMR 5 simulation is smoother than the uniform dynamic simulation since the AMR condition targets the CMEs in the domain. The time series modelled by the steady simulation has a significantly different profile before the arrival of the CMEs on June 20. Dynamic simulations better model the impact of the CMEs on Earth upon arrival. The features in AMR level 5 simulation results during the CME passage are sharper, indicating the higher resolution impacting the location of Earth in the domain.{ The profiles of the different simulations upon the first CME arrival indicate that the propagation of the CME in the dynamic solar wind yields a more similar profile to the observed data than in the steady solar wind. The modelled solar wind affects the arrival time of the CME in different simulations. The profiles are more similar in the uniform dynamic and AMR level 5 dynamic simulations, the CME arrives slightly later in the steady simulation, however, the main difference is at the arriving structure, the second shock on June 22 is more similar in all three cases. } The last CME encounters Earth around June 29, and the impact is not strong as it is not a head-on encounter. Still, nevertheless, once it passes completely, starting from July 4, we can see differences in the solar wind modelling between the steady and dynamic heliospheric models. The profile modelled by the steady heliospheric boundary driving is less similar to the observed data than in the case of dynamic heliospheric boundary driving. The comparison is consistent in all modelled variables, and the dynamic simulations show improvement compared to the steady ones. The obtained result is relatively intuitive because the data after July 1$^{st}$ is two weeks after the magnetogram timestamp used for steady simulations, and the wind is majorly outdated. There is an absence of substantial variations in the magnetic field profile during the CME passing period because the simplified, non-magnetized cone CME model is injected into the simulations.

\begin{table}[htb!]
  \caption{Run times (wall-clock time) for \texttt{Icarus 3.0} simulations.}
  \centering
   \begin{tabular}{c c  }
  \hline\hline
   Simulation & Time \\[4pt]
   \hline
Steady & 1 h 10 m\\[4pt] 
\hline
Dynamic & 1 h 4 m\\[4pt] 
\hline
Dynamic, 5AMR & 2 h 10 m\\[4pt] 
\hline
 \end{tabular}
 \tablefoot{ All the simulations were performed on six  nodes with two Xeon Gold 6240 CPUs@2.6 GHz (Cascadelake), 18 cores each, on the Genius cluster at KU Leuven.}
  \label{table:run_times_2015}

\end{table}

Table~\ref{table:run_times_2015} compares the times that simulations presented in this section require when performed on six nodes with two Xeon Gold 6240 CPUs@2.6 GHz (Cascadelake), 18 cores each, on the Genius cluster at KU Leuven. The steady simulation takes one hour and ten minutes, the dynamic simulation takes one hour and four minutes, and the simulations performed on a stretched grid with 5 AMR levels took two hours and ten minutes. There is no significant difference between the steady and dynamic simulations performed on the same computational grids. The simulation with 5 AMR levels was slower, as expected since the AMR criterion was not optimized, and it was aimed at 5 CMEs that were injected during a week, leading to big portions of refined blocks in the domain for more extended periods. The goal of the simulations with AMR levels was to demonstrate that we can apply AMR to CMEs when propagating in the dynamic solar wind, similar to the previous version.

\section{Conclusions}
\label{conclusions}
The \texttt{Icarus} heliospheric model has been upgraded to use \texttt{MPI-AMRVAC~3.0} \citep{Keppens2023} as a base MHD solver. \texttt{Icarus~3.0} was extended to dynamic solar wind modelling in the solar heliosphere. During this upgrade, built-in functionalities were favoured over manual implementation. This approach extended the co-rotating frame to the MHD module and activated in the heliospheric domain from the parameter file with the same rotation rate. The particle sampling module was generalized and adjusted to handle stretched radial grids, and the particles were injected as satellites and planets in the inner heliospheric domain. The boundary driving of the heliosphere was newly implemented using the novel boundary driving from the provided data method. The new boundary conditions were tested for the September-October period in 2019. The WSA-like coronal model was computed repeatedly for each magnetogram during the given period. The model was automated to calculate and provide inner boundary conditions with the given cadence within the indicated period. The obtained inner boundary information was stacked as a single data cube and passed to the heliospheric modelling tool. The case was chosen considering the results in \cite{Wijsen2021}, and the modelled results were compared to OMNI 1-min and Stereo-A data. In both cases, we could observe that the dynamic simulation outperforms the steady simulation for long-term simulations. The dynamic simulations also perform better in capturing a high-speed stream that was missed by the steady simulation, and the variations in number density are resolved more accurately than the {steady simulations}. 

Tables~\ref{table:run_times_2019} and ~\ref{table:run_times_2015} showed that the simulation times do not differ much when considering the cases we performed. The dynamic simulations were slightly faster in both cases than the static ones.

The standard approach in \texttt{MPI-AMRVAC} for boundary driving with data only allows setting the boundary conditions from a VTK file.  This excludes the possibility of injecting the CME into the domain. To avoid this limitation, a new boundary condition type was implemented. The new boundary condition type allows for setting the boundary conditions from the provided data source and, when necessary, modifying the boundary conditions to allow the propagation of the CME in the domain. This way, modelling the dynamic solar wind and superposing the CMEs from the inner heliospheric boundary is allowed in \texttt{Icarus}. 

Two representative cases were modelled to check the newly developed boundary conditions. The solar wind for June 2015 was modelled with five consecutive CMEs. For this study, a zero-corrected product of the GONG magnetogram was chosen instead of the standard one used before. The parameters for these CMEs were taken from \cite{Pomoell2018}. One simulation for this case was performed on a flexible computational domain with a radially stretched grid, and 5 AMR levels governed by the ``combined'' refinement criterion defined in \cite{Baratashvili2022} to demonstrate the full capabilities. In this way, the velocity and the CME tracing function divergence are refined to higher resolution in the domain, resulting in better-resolved shocks and the CME interior in the heliospheric domain. These simulations showed no significant improvement when the five consecutive CMEs passed through the location of Earth. Still, the dynamic simulations outperformed the steady simulations once the solar wind was restored. The profiles modelled by the dynamic wind are in significantly better agreement with the observed data than in the steady case.

Modelling solar wind accurately is crucial for space weather modelling as it significantly impacts CME propagation. The dynamic solar wind agrees better with the observed data than the steady solar wind driving. Therefore, the upgraded heliospheric modelling tool \texttt{Icarus} is better suited for space weather forecasting. {Our future work implies a more detailed investigation of the effect of the magnetogram processing and the dynamic boundary driving on the obtained time-dependent solar wind in the heliosphere for space weather forecasting purposes.} The dynamic boundary driving also extends the capabilities of heliospheric modelling as the solar wind does not become outdated in a few days, extending the outer heliospheric boundary farther than 2~AU to the orbit of Jupiter or beyond. {In this study, we demonstrated the upgrade to the \texttt{Icarus} model. In the follow-up studies, we intend to extend the domain to Jupiter's orbit to model a dynamic solar wind and investigate its effect near the Earth and farther out in the heliosphere. Consequently,} the accessibility to the dynamic and realistic solar wind leads to more advanced and multifaceted studies in the solar heliosphere.

\begin{acknowledgements}
This research has received funding from the European Union’s Horizon 2020 research and innovation programme under grant agreement No 870405 (EUHFORIA 2.0) and the ESA project "Heliospheric modelling techniques“ (Contract No. 4000133080/20/NL/CRS).
These results were also obtained in the framework of the projects C16/24/010  (C1 project Internal Funds KU Leuven), G0B5823N and G002523N  (FWO-Vlaanderen), 4000145223 (SIDC Data Exploitation (SIDEX2), ESA Prodex), and Belspo project B2/191/P1/SWiM.
F.B.\ acknowledges support from the FED-tWIN programme (profile Prf-2020-004, project ``ENERGY'') issued by BELSPO and from the FWO Junior Research Project G020224N granted by the Research Foundation -- Flanders (FWO).
R.K.\ acknowledges FWO projects G0B4521N and G0B9923N.
The Computational resources and services used in this work were provided by the VSC-Flemish Supercomputer Center, funded by the Research Foundation Flanders (FWO) and the Flemish Government-Department EWI.
\end{acknowledgements}

%
%

\bibliographystyle{aa}
\bibliography{bibliography}

\end{document}